\documentclass[letterpaper, 10pt, conference]{ieeeconf}  % Comment this line out
\IEEEoverridecommandlockouts                              % This command is only
\usepackage{graphics} % for pdf, bitmapped graphics files
\usepackage{epsfig} % for postscript graphics files
\usepackage{amsmath} % assumes amsmath package installed
\usepackage{amssymb}  % assumes amsmath package installed
\usepackage{xcolor}
\usepackage{mathtools}
\usepackage{cite}

\title{\LARGE \bf Observability-Blocking Control using Sparser and Regional Feedback
for Network Synchronization Processes} 

\author{Abdullah Al Maruf and Sandip Roy% <-this % stops a space
\thanks{This work was partially supported by NSF Grants 1545104 and 1635184.}% <-this % stops a space
\thanks{Authors are with School of Electrical Engineering and Computer Science,
        Washington State University, Pullman, WA 99164, USA.
        {\tt \small abdullahal.maruf@wsu.edu, sandip@wsu.edu}}%
}

\begin{document}
\maketitle
%\thispagestyle{empty}
%\pagestyle{empty}

%%%%%%%%%%%%%%%%%%%%%%%%%%%%%%%%%%%%%%%%%%%%%%%%%%%%%%%%%%%%%%%%%%%%%%%%%%%%%%%%
\begin{abstract}

The design of feedback control systems to block observability in a network synchronization model, i.e. to  make the dynamics unobservable from measurements at a subset of the network's nodes, is studied. First, a general design algorithm is presented for blocking observability at any specified group of $m$ nodes, by applying state feedback controls at $m+2$ specified actuation nodes.  The algorithm is based on a method for eigenstructure assignment, which allows surgical modification of particular eigenvectors to block observability while preserving the remaining open-loop eigenstructure. Next, the topological structure of the network is exploited to reduce the number of controllers required for blocking observability; the result is based on blocking observability on the nodes associated with a vertex-cutset separating the actuation and measurement locations. Also, the design is modified to encompass regional feedback controls, which only use data from a subset of accessible nodes. The regional feedback design does not maintain the open-loop eigenstructure, but can be guaranteed to preserve stability via a time-scale argument. The results are illustrated with numerical examples.

\end{abstract}

\section{Introduction}

Assessing controllability and, dually, observability of complex dynamical networks has been a significant research focus of the network-controls community in the last few years \cite{control1,control2,control3,control4,control5,control6,control7,control8,control9,control10,control11,control12,control13}. A main outcome of this effort is a set of graph-theoretic necessary or sufficient conditions for observability and controllability, for canonical linear network models (e.g. synchronization models) \cite{control2,control8,control10}.  This core graph-theoretic analysis has been extended in several directions, including toward evaluating Gramian-based metrics for control energy and state-estimation fidelity \cite{control3, control6,control8}, studying input-output notions including output controllability and transfer-function zeros \cite{control5,control11}, and addressing sensor/actuator placement \cite{control7,control12}.  Analyses have also been undertaken for increasingly sophisticated linear models \cite{control9}, and recently for some simple classes of nonlinear models \cite{control13}.  These studies have been motivated by a number of applications, including cyber-threat assessment of multi-agent systems, infrastructure control and monitoring, and privacy-preserving design of network algorithms.

The  studies on observability and controllability of network processes have largely focused on the relationship between a network's native topology and its input-output behavior. A key further question of interest is whether a network's control systems can be designed to facilitate or block observability/controllability from particular network locations, while preserving overall performance.  Such design problems often arise when multiple control authorities or stakeholders have access to the network's dynamics, and have cross-cutting goals in modulating the dynamics (whether cooperative or competitive).  For instance, given increasing concern about cyber-attacks in the power grid, there is interest in designing wide-area control systems that not only damp oscillations but reduce the observability or controllability of adversaries \cite{new_1}.  Likewise, control initiatives in the air transportation system can be designed to improve or prevent estimation of flow dynamics \cite{new_2}, and control of water distribution systems to prevent malfeasance is of interest \cite{seanpap}. At a different scale, controller designs for multi-vehicle systems may need to consider security from intruders that can probe a subset of vehicles \cite{new_3}. Therefore, there is a motivation to study the design of control systems in networks, to either facilitate or prevent state estimation. Here, we pursue a study in this direction, focused on designing controls at some network nodes to make the network's dynamics unobservable at remote locations.

Specifically, in this study, we consider a  linear network synchronization model \cite{cons1,cons2,cons3,cons4,cons5} which can be actuated at a sparse set of network nodes. Our goal is to determine whether regional state feedback controls applied at these nodes can be used to enforce unobservability to monitoring at a set of remote nodes, while maintaining dynamical properties of the network. We study the design of such {\em observability-blocking controllers}, by applying and enhancing a method for eigenstructure assignment \cite{jointeig,jointeig2,surgeig}. The main contributions of the study are as follows:   

\begin{itemize}
\item A general state-feedback controller design algorithm is obtained, which blocks observability at a set of measurement nodes through assignment of a (or few) closed-loop eigenvector(s). Under broad conditions, the other open-loop eigenvectors and eigenvalues can be preserved.

\item A topology-exploiting scheme which requires fewer actuation nodes to block observability is developed. It is shown that the required number of actuators depends on the cardinality of the vertex-cutset separating vertices corresponding to the actuation nodes and measurement nodes.

\item The design scheme is further modified to accommodate regional feedback where only the states from a region or partition is used. In this case, the observability-blocking control scheme modifies the eigenstructure of the open-loop system, but stability can still be guaranteed via a time-scale separation-based design.

\end{itemize}

The research described here connects to a wide literature on controller design for built network processes, such as stabilizer design to improve damping of inter-area oscillations in the power grid  \cite{new_7}, mitigation of infection spreads \cite{control15}, and pinning control of synchronization processes \cite{control16}. However, these various studies have been primarily focused on shaping the internal dynamics of network processes, while our interest here lies in shaping external or channel properties. We point the reader toward  some preliminary work on shaping input-output dynamics of networks \cite{control12,control13}, which is focused particularly on invariant zeros of network channels.   The research described here also connects to the classical body of work on shaping the spectrum of linear systems (eigenvalues and eigenvectors) via feedback \cite{eig-contl_lin,jointeig_para,jointeig}.  Relative to this literature, the contribution of our work lies in the development of pruned designs which exploit a graph structure overlaid on the linear dynamics. 

The paper is organized as follows. In Section II, we formulate the observability-blocking controller design problem. In Section III, eigenstructure assignment via linear state feedback control, which is the basis for our approach, is briefly reviewed  \cite{jointeig,surgeig}. The main results (design algorithms, proofs) are given in Section IV. Examples are given in Section V, and brief concluding remarks are included in Section VI. Some preliminary results in this direction were presented in \cite{new_5}. Relative to \cite{new_5}, corrections and extensions of the design algorithms have been undertaken, the regional feedback design has been developed, new examples are included, and the motivation of the work has been updated.

\section{Problem Formulation}

We consider a standard model for network synchronization  \cite{cons1,cons2,cons3,cons4,cons5}, which has been enhanced to represent: 1) actuation nodes where feedback controls can be applied by a system operator, and 2) measurements or outputs available to a stakeholder (e.g. an adversary). Our primary objective is to design feedback controllers at the actuation nodes such that the dynamics are unobservable to the measurement nodes.
We focus on synchronization processes as a representative class, because they arise in many settings where estimation-prevention may be important (e.g. robotic teams, water-distribution systems).

Formally, the synchronization model is defined on a weighted digraph $\mathcal{G}(\mathcal{V},\mathcal{E}:\mathcal{W})$. Here, the vertex set $\mathcal{V}$ contains $n$ vertices labeled as $1,2, \hdots, n$.  Each directed edge in $\mathcal{E}$ is specified as an ordered pair of vertices, i.e. $(i,k)$ indicates an edge from vertex $i$ to vertex $k$.  The weight of each edge $(i,k) \in \mathcal{E}$ is denoted by $w_{i k}$, and is assumed to be positive.  For this study, we assume the network graph to be strongly connected. 
The synchronization dynamics is specified by the (asymmetric) Laplacian or diffusion matrix $\mathbf{L}$ of the graph. Specifically, $\mathbf{L}$ is an $n \times n$ matrix whose entries are as follows: each off-diagonal entry $L_{k i}$ is equal to $-w_{ik}$ for all $(i,k) \in \mathcal{E}$, otherwise $0$; each diagonal entry $L_{kk}$ is equal to $-\sum_{i=1, i \neq k}^{n} L_{ki}$.  

The synchronization model comprises a network with $n$ nodes labeled $1, 2, \hdots, n$, which correspond to the graph vertices.  Each node $i$ has associated with it a scalar state $x_i(t)$.  The network state is defined as $\mathbf{x}(t)= [x_1(t) ~~ x_2(t) ~ \cdots ~ x_n(t)]^T$. Actuation can be provided at a set of $q$ nodes $\{r_1, r_2, \hdots, r_q\}$, which we call actuation nodes (respectively, actuation vertices in the graph). Also, the states of $m$ measurement nodes (measurement vertices in the graph) $\{s_1, s_2, \hdots, s_m\}$ are observed. We stress that the actuation and measurement nodes may overlap. The model dynamics are then given by:
\begin{eqnarray}
\mathbf{\dot{x}} &=& -\mathbf{L}\mathbf{x}+\mathbf{B}\mathbf{u}, \\
\mathbf{y} &=& \mathbf{C}\mathbf{x}
\end{eqnarray}
where $\mathbf{u} \in \mathbb{R}^q$ is a vector containing the input signals at the actuation nodes, $\mathbf{y} \in \mathbb{R}^m$ is a vector containing the observations at the measurement nodes, $\mathbf{B}=[\mathbf{e}_{r_1} ~ \mathbf{e}_{r_2} \cdots ~\mathbf{e}_{r_q}]$, $\mathbf{C}=[\mathbf{e}_{s_1} ~ \mathbf{e}_{s_2} \cdots ~\mathbf{e}_{s_m}]^T$, and $\mathbf{e}_i$ is a 0-1 indicator vector in $\mathbb{R}^n$ with only the $i$th entry equal to $1$. We assume the pair ($-\mathbf{L},\mathbf{B})$ is controllable.

In this study, we consider the design of linear feedback controllers at the actuation nodes, to modulate the observability of the network dynamics with respect to the measurements ${\bf y}$.  As a nominal case, a state feedback control scheme is considered.  The controller at each node $r_i$, $i=1, 2, \hdots, q$, is thus specified as
$u_{r_i}=-\mathbf{k}_{r_i}^T\mathbf{x}$, where $\mathbf{k}_{r_i}$ is the control gain.  Assembling
the state feedback models for each actuation node yields:
\begin{equation}
\mathbf{u}=-\mathbf{F}\mathbf{x}
\end{equation}
where $\mathbf{F}=[\mathbf{k}_{r_1} ~\mathbf{k}_{r_2} \cdots \mathbf{k}_{r_q}]^T$.
Upon application of the feedback control, the closed-loop dynamics become:
\begin{eqnarray}
\mathbf{\dot{x}}
&=& -(\mathbf{L}+\mathbf{B}\mathbf{F})~\mathbf{x}, \\
\mathbf{y} &=& \mathbf{C}\mathbf{x}. \label{eq:main}
\end{eqnarray}

Our first goal in this study is to design the state-feedback controller, defined by the gain matrix $\mathbf{F}$, to make the closed-loop model unobservable from the measurements, while preserving other dynamical properties of the open-loop system. That is, we seek to enforce that the pair $(\mathbf{C},-(\mathbf{L}+\mathbf{BF}))$ is unobservable, while maintaining as much of the open-loop eigenstructure as possible. 
Second, we seek to develop sparse controller designs which
can be used to block observability (i.e. make the pair $(\mathbf{C},-(\mathbf{L}+\mathbf{BF}))$ unobservable) using a lesser number of actuation nodes, by exploiting the graph topology of the network-synchronization model. Third, we study whether regional feedback controls (which only use state measurements in a partition of the network) can be used to block observability, while still preserving stability.

\section{Preliminaries}

Our proposed method for designing observability-blocking controllers draws on and extends techniques for eigenstructure assignment \cite{surgeig,jointeig}. The classical study in this direction by Moore et al \cite{jointeig} presents  necessary and sufficient conditions for full eigenstructure assignment when closed-loop eigenvalues are distinct, while several follow-up studies address the repeated-eigenvalue case and give alternative perspectives on assignment \cite{jointeig2,jointeig_para}; these methods have found application in several domains such as aircraft control and malicious control of the power grid \cite{jointeig_app1,jointeig_app2,jointeig_app3}.   Our recent note \cite{surgeig} has clarified that the classical eigenstructure assignment method can be adapted for surgical eigenstructure assigment -- i.e. placement of all eigenvalues and a subset of eigenvectors -- subject to conditions only on the dictated eigenvectors. Key results on surgical and full eigenstructure assignment are summarized in a single proposition here. 

To present these results on eigenstructure assignment, we require some further notations.  Formally, let us consider the closed-loop dynamics of a controllable linear system with an applied state feedback controller: $\underline{\mathbf{\dot{x}}}= (\underline{\mathbf{A}}+\underline{\mathbf{B}}~ \underline{\mathbf{F}})~\underline{\mathbf{x}}$ where $\underline{\mathbf{x}} \in \mathbb{R}^n$, $\underline{\mathbf{A}} \in \mathbb{R}^{n \times n}$, $\underline{\mathbf{B}} \in \mathbb{R}^{n \times q}$ and $\underline{\mathbf{F}} \in \mathbb{R}^{q \times n}$ denote the state, open-loop state matrix, input matrix and gain matrix, respectively. Then, for any given complex $\underline{\lambda}$, we construct a matrix  $\mathbf{N}(\underline{\lambda}) = [\mathbf{N}_1(\underline{\lambda})^T ~~\mathbf{N}_2(\underline{\lambda})^T]^T$ where $\mathbf{N}(\underline{\lambda}) \in \mathbb{C}^{(n+q) \times q}, \mathbf{N}_1(\underline{\lambda}) \in \mathbb{C}^{n \times q}, \mathbf{N}_2(\underline{\lambda}) \in \mathbb{C}^{q \times q}$, to have the following properties: the columns of $\mathbf{N}(\underline{\lambda})$ are linearly independent and span the null space of $\mathbf{S}(\underline{\lambda})= [(\underline{\mathbf{A}}-\underline{\lambda}~\mathbf{I}_n)~~ \underline{\mathbf{B}}]$. In this notation, the main results developed in \cite{jointeig} and \cite{surgeig} can be summarized as follows:

\medskip

\textbf{Proposition 1}: {\em Assume that $\{\underline{\lambda}_1,\underline{\lambda}_2,\hdots, \underline{\lambda}_n\}$ is any self-conjugate set of distinct complex numbers and the pair ($\underline{\mathbf{A}},\underline{\mathbf{B}}$) is controllable. Then a real valued state feedback controller gain matrix $\underline{\mathbf{F}}$ can be designed to: 1) place any $r$ closed-loop eigenvalues at $\underline{L}_1$ $=\{\underline{\lambda}_1, \underline{\lambda}_2, \hdots, \underline{\lambda}_r\}$ and their corresponding eigenvectors at any $\underline{V}_1$ $= \{ \underline{\mathbf{v}}_1, \underline{\mathbf{v}}_2, \hdots, \underline{\mathbf{v}}_r \}$, and 2) place the remaining $n-r$ closed-loop eigenvalues at $\underline{L}_2$ $=\{\underline{\lambda}_{r+1}, \underline{\lambda}_{r+2}, \hdots, \underline{\lambda}_n\}$ provided that the following conditions hold: (i) $\underline{V}_1$ is a set of linearly independent vectors in $\mathbb{C}^n$, (ii) $\underline{\mathbf{v}}_i= \bar{\underline{\mathbf{v}}}_k$ whenever $\underline{\lambda}_i = \bar{\underline{\lambda}}_k$ where $i, k \in \{1, 2, \hdots, r\}$, (iii) $\underline{\mathbf{v}}_i \in$ column space of $\mathbf{N_1}(\underline{\lambda}_i)$ for all $i \in \{1, 2, \hdots, r\}$, (iv) $\underline{L}_2$ is a self-conjugate set. Furthermore, for this feedback design, the closed-loop eigenvectors $\underline{V}_2$ $= \{ \underline{\mathbf{v}}_{r+1}, \underline{\mathbf{v}}_{r+2}, \hdots, \underline{\mathbf{v}}_n \}$ corresponding to the eigenvalues in $\underline{L}_2$ satisfy:  $\underline{\mathbf{v}}_i \in$ column space of $\mathbf{N_1}(\underline{\lambda}_i)$ for all $i \in \{r+1, r+2, \hdots, n\}$.}

\medskip

Generalizations of the result for the repeated eigenvalue case have been presented  in \cite{jointeig2, surgeig}; details are omitted here to simplify the presentation. 

\section {Main Results}

The design of observability-blocking controllers is undertaken in three steps. First, we develop a general algorithm for blocking observability  via state feedback control, when the number of actuation nodes $q$ exceeds the number of measurement nodes $m$ by two ($q=m+2$), see Section IV.A; the algorithm is general in the sense that it does not depend on the network graph or even the Laplacian form of the state matrix.  Then, we demonstrate that a smaller set of actuation nodes can be used, if the network's graph has a low-cardinality separating cutset between the actuation vertices and measurement vertices. In both cases, the controllers entirely preserve the open-loop dynamics, except perhaps a small set of eigenvectors. Lastly, we show that regional feedback controllers, which only use states from a region or partition of the network, are also able to block observability. These regional controllers distort the open-loop eigenstructure in general, but can be designed to guarantee stability.

\subsection{General Design of Observability-Blocking Controllers}

We present an algorithm for designing a state-feedback controller which makes the the pair $(\mathbf{C},-(\mathbf{L}+\mathbf{BF}))$ unobservable. The design is based on the eigenstructure assignment method reviewed above. Specifically, the algorithm places one eigenvector of the closed-loop state matrix to block observability, while aiming to keep all eigenvalues and most of the remaining eigenvectors unchanged. Precisely, the algorithm places one eigenvector $\mathbf{v}$ of  $-(\mathbf{L}+\mathbf{BF})$ to satisfy $\mathbf{Cv}= \mathbf{0}$ (while preserving the remainder of the eigenstructure), whereupon unobservability follows from the Popov-Belevitch-Hautus or PBH test \cite{eig-contl_lin}.  

In the ensuing development, for notational convenience and without loss of generality, the last $m$ nodes (nodes $n-m+1, \hdots, n$) are assumed to be the measurement nodes. Hence, $\mathbf{C}=[\mathbf{e}_{n-m+1} ~\cdots ~\mathbf{e}_n]^T$, and the the pair $(\mathbf{C},-(\mathbf{L}+\mathbf{BF}))$ is unobservable from the PBH test if and only if $(\mathbf{L}+\mathbf{BF})$ has an eigenvector whose final $m$ entries are zero.

We first present the design algorithm for the case where the open-loop eigenvalues are distinct and real, and then use this result to obtain algorithms for more general settings. For the case with distinct and real eigenvalues, we label the eigenvalues of $\mathbf{L}$ as $\lambda_1, \lambda_2, \hdots, \lambda_n$, and note that corresponding eigenvectors $\mathbf{v}_1, \mathbf{v}_2, \hdots, \mathbf{v}_n$ can be found which are linearly independent and span $\mathbb{R}^n$. We also define the sets $L_0$ and $V_0$ which contain the open-loop eigenvalues and eigenvectors respectively. 

The design algorithm is based on selecting any one of the eigenvalues of $\mathbf{L}$ and its associated eigenvector, say $\lambda_p$ and $\mathbf{v}_p$ where $p \in \{1, 2, \hdots, n \}$. The goal of the design is to construct a feedback control which uses $q=m+1$ actuation nodes, such that the eigenvector $\mathbf{v}_p$ is modified to a vector $\mathbf{\hat{v}}_p$ whose entries corresponding to the measurement nodes are all zeros, while most of the remaining eigenvectors and all the eigenvalues are maintained. In this way, the mode $(-\lambda_p)$ of the system dynamics is made unobservable. This can be done according to the following steps:

\noindent \textbf{Algorithm 1:}
\begin{enumerate}

\item Select one eigenvalue $\lambda_p$ of $\mathbf{L}$ and its associated eigenvector $\mathbf{v}_p$,  where $p \in \{1, 2, \hdots, n\}$.

\item Compute a matrix $\mathbf{N}(\lambda_p) \in \mathbb{R}^{(n+q) \times q}$, whose columns are linearly independent and span the null space of $\mathbf{S}(\lambda_p)=[(\mathbf{L}-\lambda_p\mathbf{I}_n)~~ \mathbf{B}]$.  Then partition $\mathbf{N}(\lambda_p)$ as $\mathbf{N}(\lambda_p)=[\mathbf{N}_1(\lambda_p)^T ~ \mathbf{N}_2(\lambda_p)^T]^T$, where  $\mathbf{N}_1(\lambda_p) \in \mathbb{R}^{n \times q}$ and  $\mathbf{N}_2(\lambda_p) \in \mathbb{R}^{q \times q}$, as described in Section III. Therefore $\mathbf{N}_1(\lambda_p)$ and $\mathbf{N}_2(\lambda_p)$ satisfy:
\begin{eqnarray}
[(\mathbf{L}-\lambda_p~\mathbf{I}_n)~~ \mathbf{B}] \left[\begin{array}{c}
                                         \mathbf{N}_1(\lambda_p)   \\
                                         \mathbf{N}_2(\lambda_p)
                                          \end{array}\right]  = \mathbf{0}.  \label{eq4}
\end{eqnarray}

\item Partition $\mathbf{N}_1(\lambda_p)$ as
$\mathbf{N}_1(\lambda_p)=[\mathbf{N}_3(\lambda_p)^T ~ \mathbf{N}_4(\lambda_p)^T]^T$, where  $\mathbf{N}_3(\lambda_p) \in \mathbb{R}^{(n-m) \times q}$ and  $\mathbf{N}_4(\lambda_p) \in \mathbb{R}^{m \times q}$.  Then find a vector $\mathbf{h}_p \neq \mathbf{0}$ which lies in the null space of $\mathbf{N}_4(\lambda_p)$, i.e. which satisfies:  
\begin{eqnarray}
\mathbf{N}_4(\lambda_p)~ \mathbf{h}_p &=& \mathbf{0}. \label{eq2}
\end{eqnarray}

\item Compute the vectors $\mathbf{\hat{v}_p}$ and $\mathbf{z}_p$ as:
\begin{eqnarray}
\mathbf{\hat{v}}_p &=& \mathbf{N}_1(\lambda_p)~ \mathbf{h}_p \label{eq1} \\
\mathbf{z}_p &=& \mathbf{N}_2(\lambda_p)~ \mathbf{h}_p. \label{eq5}
\end{eqnarray}

\item Form the open-loop modal matrix $\mathbf{V}_0 \in \mathbb{R}^{n \times n}$ (i.e. $\mathbf{V}_0=[ \mathbf{v}_1 \mathbf{v}_2 \cdots \mathbf{v}_n]$), and also a zero matrix $\mathbf{Z}_0$ in  $\mathbb{R}^{q \times n}$. Then check whether the $n$ vectors in $\hat{V}_0 = V_0 \cup \{\hat{\mathbf{v}}_p \} \backslash \{ \mathbf{v}_p \}$ are linearly independent. If yes, then follow Step 6 below to construct the closed-loop modal matrix $\mathbf{V}$ and associated $\mathbf{Z}$ matrix. Otherwise, skip Step 6, and follow steps 7 through 9 to construct the $\mathbf{V}$ and $\mathbf{Z}$ matrices. 

\item Construct the matrix $\mathbf{V} \in \mathbb{R}^{n \times n}$ from $\mathbf{V}_0$ by replacing the column containing $\mathbf{v}_p$ with $\mathbf{\hat{v}_p}$. Similarly, construct the matrix $\mathbf{Z} \in \mathbb{R}^{q \times n}$ from $\mathbf{Z}_0$ by replacing the corresponding column with $\mathbf{z}_p$.
Therefore, $\mathbf{V}=[ \mathbf{v}_1 \cdots \mathbf{v}_{p-1}~ \mathbf{\hat{v}_p}~\mathbf{v}_{p+1}\cdots \mathbf{v}_n]$ and $\mathbf{Z}= [\mathbf{0} \cdots \mathbf{0} ~ \mathbf{z}_p~ \mathbf{0} \cdots \mathbf{0}]$. Then jump to Step 10.

\item Choose another eigenvalue and eigenvector pair $(\lambda_k,\mathbf{v}_k)$ such that the $n-1$ vectors in $V_1=V_0 \cup \{\hat{\mathbf{v}}_p \} \backslash \{\mathbf{v}_p, \mathbf{v}_k \}$ are linearly independent.

\item Then find a vector $\hat{\mathbf{v}}_k$ in the column space of $\mathbf{N}_1(\lambda_k)$ such that the $n$ vectors in $\hat{V}=V_0 \cup \{\hat{\mathbf{v}}_p, \hat{\mathbf{v}}_k \} \backslash \{ \mathbf{v}_p, \mathbf{v}_k \}$ are linearly independent. Then compute $\mathbf{z}_k$ as:
\begin{equation}
\mathbf{z}_k = \mathbf{N}_2(\lambda_k)~ \mathbf{h}_k ~~\mbox{such that}~~
\mathbf{\hat{v}}_k = \mathbf{N}_1(\lambda_k)~ \mathbf{h}_k. \label{eq5b} \\
\end{equation}

\item Construct the matrix $\mathbf{V} \in \mathbb{R}^{n \times n}$ from $\mathbf{V}_0$ by replacing the columns containing $\mathbf{v}_p$ and  $\mathbf{v}_k$ with $\mathbf{\hat{v}}_p$ and $\mathbf{\hat{v}}_k$ respectively. Similarly, construct the matrix $\mathbf{Z} \in \mathbb{R}^{q \times n}$ from $\mathbf{Z}_0$ by replacing the corresponding columns with $\mathbf{z}_p$ and $\mathbf{z}_k$. Hence $\mathbf{V}=[ \mathbf{v}_1 \cdots  \mathbf{v}_{p-1}~ \mathbf{\hat{v}}_p ~ \mathbf{v}_{p+1} \cdots$ $\mathbf{v}_{k-1}~\mathbf{\hat{v}}_k~\mathbf{v}_{k+1} \cdots \mathbf{v}_n]$ and $\mathbf{Z}= [\mathbf{0} \cdots \mathbf{0}~\mathbf{z}_p~\mathbf{0}$ $\cdots \mathbf{0}~\mathbf{z}_k ~ \mathbf{0}  \cdots \mathbf{0}]$ when $p<k$. 

\item Finally the gain matrix $\mathbf{F}$ for the observability-blocking controller is uniquely obtained by (\ref{eq3a}):
\begin{eqnarray}
 \mathbf{F} = \mathbf{Z}~\mathbf{V}^{-1}.  \label{eq3a}
\end{eqnarray}
\end{enumerate}

As demonstrated in the following theorem, the controller designed according to the Algorithm 1 guarantees that the pair $(\mathbf{C},-(\mathbf{L}+\mathbf{BF})$ is unobservable. Furthermore, it preserves all open-loop eigenvalues, and all but one ($\mathbf{v}_p$) or possibly two ($\mathbf{v}_p$, $\mathbf{v}_k$) eigenvectors.

\medskip

\noindent \textbf{Theorem 1:}  \textit{Consider the network synchronization model. Assume that: 1) the eigenvalues of $\mathbf{L}$ are distinct and real, 2) the number of actuation nodes is one more than the number of measurement nodes ($q=m+1$), and 3) the pair ($-\mathbf{L},\mathbf{B})$ is controllable. Then the gain matrix $\mathbf{F}$ of the state feedback controller obtained using Algorithm 1 blocks the observability of the model, specifically the pair $(\mathbf{C},-(\mathbf{L}+\mathbf{BF}))$ has an unobservable mode at $(-\lambda_p)$. Furthermore, if the set $\hat{V}_0$ obtained in Step 5 contains $n$ linearly independent vectors, then all  eigenvalues and all except one eigenvector ($\mathbf{v}_p$) of the open-loop model are maintained in the closed-loop model. Otherwise, all eigenvalues and all except two eigenvectors ($\mathbf{v}_p$, $\mathbf{v}_k$) of the open-loop model  are maintained in the closed-loop model.}

\medskip

\noindent \textbf{Proof:}

First note that $\mathbf{N}_4(\lambda_p) \in \mathbb{R}^{m \times q}$, as defined in Step 3 of Algorithm 1, is a rank deficient matrix since $q=m+1$. Therefore, a nonzero vector $\mathbf{h}_p$ satisfying (\ref{eq2}) is guaranteed to exist. Now consider two cases separately for the proof: Case 1 where $\hat{V}_0$ is a set of $n$ linearly independent vectors, and Case 2 where $\hat{V}_0$ is not a set of $n$ linearly independent vectors.

For the Case 1, notice that $\mathbf{V}$ defined in Step 6 is invertible. Then consider  the gain matrix $\mathbf{F}$ obtained in Step 10 in this case. From (\ref{eq3a}), it follows that $\mathbf{F {v}}_i = \mathbf{0}$ $~\forall i \in \{1, \cdots, n \} \backslash \{p\}$, therefore all the open-loop eigenvalue and eigenvector pairs beside ($\lambda_p,\mathbf{v}_p$) are maintained in the closed-loop model. Now we will show that ($\lambda_p,\hat{\mathbf{v}}_p$) is an eigenvalue and eigenvector pair of $\mathbf{L+BF}$, using similar arguments to \cite{jointeig}. By multiplying (\ref{eq4}) with $\mathbf{h}_p$ and using (\ref{eq1}), (\ref{eq5}) and (\ref{eq3a}) we easily find that $(\mathbf{L}+\mathbf{BF})\mathbf{\hat{v}}_p=\lambda_p\mathbf{\hat{v}}_p$. Hence, all the eigenvalues and eigenvectors of $\mathbf{L}$ except $\mathbf{v}_p$ is maintained in $\mathbf{L+BF}$ for Case 1 where $\mathbf{\hat{v}}_p$ is the new closed-loop eigenvector.

Now consider the Case 2. For this case, first note that $\mathbf{N}_1(\lambda_p)$ is full rank because $\mathbf{B}$ is full rank \cite{jointeig,new_5}. Therefore $\hat{\mathbf{v}}_p$ obtained from (\ref{eq1}) in Step 4 is non-zero. Since
$\hat{\mathbf{v}}_p$ is a linear combination of the open-loop eigenvectors other than $\mathbf{v}_p$, at least one eigenvector $\mathbf{v}_k$ can be found in Step 7 such that $V_1$ is a set of $(n-1)$ linearly independent vectors. Now consider $L_1= L_0 \backslash \{\lambda_k\}$ and $L_2=\{\lambda_k\}$. Note that $L_1$, $V_1$ and $L_2$ satisfy all the conditions given in Proposition 1. Therefore by applying Proposition 1, we can guarantee the existence of $\hat{\mathbf{v}}_k$ in Step 8 such that $\hat{\mathbf{v}}_k$ is in the column space of $\mathbf{N}_1(\lambda_k)$ and $\hat{V}=V_0 \cup \{\hat{\mathbf{v}}_p, \hat{\mathbf{v}}_k \} \backslash \{ \mathbf{v}_p, \mathbf{v}_k \}$ is a set of $n$ linearly independent vectors. Hence in the Case 2, $\mathbf{V}$ defined in Step 9 is invertible. Using arguments similar to those for Case 1, we can easily show that all the eigenvalues and eigenvectors of $\mathbf{L}$ except $\mathbf{v}_p$ and $\mathbf{v}_k$ are maintained in $\mathbf{L+BF}$ for Case 2 where $\mathbf{\hat{v}}_p$ and $\mathbf{\hat{v}}_k$ are the new closed-loop eigenvectors.

Finally note that the last $m$ entries of $\mathbf{\hat{v}}_p$ are zero according to (\ref{eq2}), hence the mode $(-\lambda_p)$ of the closed-loop system is unobservable, which completes our proof. \hfill $\blacksquare$

\medskip

In the above algorithm, the check on linear independence of the vectors contained in $\hat{V}_0$ in Step 5 is needed, as it dictates the construction of closed-loop modal matrix $\mathbf{V}$. When $\hat{V}_0$ is not a set of linearly independent vectors, $\mathbf{V}$ as defined in Step 6 is not invertible. In this case, we follow Steps 7, 8 and 9 instead to construct an invertible $\mathbf{V}$ in order to obtain the gain matrix $\mathbf{F}$. We note that our preliminary work in this direction \cite{new_5} had an inaccuracy, in that this condition on $\hat{V}_0$ and resulting additional steps were not indicated. The following example demonstrates this situation can occur, i.e. $\hat{V}_0$ may not be a set of $n$ linearly independent vectors. However, we note that the situation arises only for specially constructed examples, and hence the additional steps are unnecessary in typical cases. 

{\em Example:} Consider a network model with parameters $\mathbf{L}=[2, 0,$ $-1, -1;$ $0, 3, -3,$ $0; -1, -1,$ $5, -3; -1,$ $0, -1, 2]$, $\mathbf{B}= [1, 0, 0;$ $0, 1, 0;$ $0, 0, 1;$  $0, 0, 0]$ and $\mathbf{C}=[0, 0, 1, 0;$ $0, 0, 0, 1]$. The Laplacian has eigenvalues $0, 2.4384,3$ and $6.5616$. If we choose $\lambda_p=3$, the vectors in $\hat{V}_0$ are linearly dependent. This happens because $\mathbf{\hat{v}}_p$ obtained from (\ref{eq1}) is in the range space of the eigenvectors associated with the eigenvalues at $0, 2.4384$ and $6.5616$. Now according to our algorithm we can choose $\lambda_k=6.5616$ in Step 7. Then we choose $\mathbf{\hat{v}}_k=[-1.0569; 0.6022;$ $-0.4802; 0.3370]$ from the null space of $\mathbf{N}_1(6.5616)$ such that $\hat{V}$ is a set of four linearly independent vectors. This choice yields $\mathbf{F}=[3.5616,$ $0,$ $0,$ $-3.5616;$ $-0.5051, 0,$ $0, 0.5051;$ $2.4174, 1,$ $ -3.5616, 0.1442]$. This observability-blocking controller maintains the open-loop eigenstructure except eigenvectors $\mathbf{v}_p$ and $\mathbf{v}_k$, as stated in the Theorem 1.

Next, we modify Algorithm 1 to address the case that $\mathbf{L}$ has complex eigenvalues.  The modified algorithm requires $q=m+2$ actuation nodes to be used, in the case where the eigenvectors associated with a complex-conjugate pair are modified to achieve unobservability. Here is the algorithm:

\noindent \textbf{Algorithm 2:} 

\begin{enumerate}

\item [1)]  Select one eigenvalue $\lambda_p$ of $\mathbf{L}$ and its associated eigenvector $\mathbf{v}_p$, where $p \in \{1, 2, \hdots, n\}$. Then if $\lambda_p$ is real, follow the steps under Sub-Algorithm 1 to obtain observability-blocking controller. Otherwise follow the steps under Sub-Algorithm 2.
\end{enumerate}

\textbf{Sub-Algorithm 1: }

\begin{enumerate}

\item [2)-4)]  Steps 2 through 4 are exactly the same as Algorithm 1.

\item [5)] Form the open-loop modal matrix $\mathbf{V}_0 \in \mathbb{R}^{n \times n}$ and the zero matrix $\mathbf{Z}_0$ in  $\mathbb{R}^{q \times n}$. Then check whether the $n$ vectors in $\hat{V}_0 = V_0 \cup \{\hat{\mathbf{v}}_p \} \backslash \{ \mathbf{v}_p \}$ are linearly independent. If yes, then follow Step 6 below to construct the closed-loop modal matrix $\mathbf{V}$ and associated $\mathbf{Z}$ matrix. Otherwise, skip Step 6, and follow steps 7 through 9 to construct the $\mathbf{V}$ and $\mathbf{Z}$ matrices. 

\item [6)] Construct the matrix $\mathbf{V} \in \mathbb{R}^{n \times n}$ from $\mathbf{V}_0$ by replacing the column containing $\mathbf{v}_p$ with $\mathbf{\hat{v}_p}$. Similarly, construct the matrix $\mathbf{Z} \in \mathbb{R}^{q \times n}$ from $\mathbf{Z}_0$ by replacing the corresponding column with $\mathbf{z}_p$. Then jump to Step 10.

\item [7)] Find the largest-cardinality subset $V_1$ of $\hat{V}_0$ such that $\hat{\mathbf{v}}_p \in V_1$ and $V_1$ is a self-conjugate set of linearly independent vectors. 

\item [8)] Find $\hat{\mathbf{v}}_k$ in the column space of $\mathbf{N}_1(\lambda_k)$ for all $\mathbf{v}_k \in \hat{V}_0 \backslash V_1$ such that the set $\hat{V}= V_1 \cup \{\hat{\mathbf{v}}_k | \mathbf{v}_k \in \hat{V}_0 \backslash V_1 \}$ is a self conjugate set of $n$ linearly independent vectors. While doing so, maintain $\hat{\mathbf{v}}_{k_2}= \bar{\hat{\mathbf{v}}}_{k_1}$ whenever $\mathbf{v}_{k_2}= \bar{\mathbf{v}}_{k_1}$ and $\mathbf{v}_{k_1}, \mathbf{v}_{k_2} \in \hat{V}_0 \backslash V_1$. Next, find the corresponding $\mathbf{z}_k$ based on (\ref{eq5b})  for all $\mathbf{v}_k \in \hat{V}_0 \backslash V_1$ (i.e. $\mathbf{z}_k = \mathbf{N}_2(\lambda_k)~ \mathbf{h}_k$, where $\mathbf{h}_k$ solves
$\mathbf{\hat{v}}_k = \mathbf{N}_1(\lambda_k)~ \mathbf{h}_k$). Note that $\mathbf{z}_{k_2}= \bar{\mathbf{z}}_{k_1}$ whenever $\mathbf{v}_{k_2}= \bar{\mathbf{v}}_{k_1}$. 

\item [9)] Construct $\mathbf{V}$ from $\mathbf{V}_0$ by replacing the columns containing $\mathbf{v}_p$ and all $\mathbf{v}_k\in \hat{V}_0 \backslash V_1$ with $\hat{\mathbf{v}}_p$ and corresponding $\hat{\mathbf{v}}_k$ respectively. In the same manner construct $\mathbf{Z}$ from $\mathbf{Z}_0$ by replacing the corresponding columns of $\mathbf{Z}_0$ with $\mathbf{z}_p$ and all $\mathbf{z}_k$ obtained in Step 8 respectively. 

\item [10)] Finally, compute the gain matrix $\mathbf{F}$ using (\ref{eq3a}).

\end{enumerate}

\textbf{Sub-Algorithm 2:} 

\begin{enumerate} 

\item [2)-4)] Steps 2 through 4 remain exactly the same as Algorithm 1. Additionally in Step 4 obtain $\bar{\hat{\mathbf{v}}}_p$ and associated $\bar{\mathbf{z}}_{p}$ by taking complex conjugates of $\mathbf{\hat{v}}_{p}$ and $\mathbf{z}_{p}$, respectively. 

\item [5)] Form the open-loop modal matrix $\mathbf{V}_0 \in \mathbb{R}^{n \times n}$ and the zero matrix $\mathbf{Z}_0$ in  $\mathbb{R}^{q \times n}$. Then check whether $\hat{V}_0 = V_0 \cup \{\hat{\mathbf{v}}_p, \bar{\hat{\mathbf{v}}}_p\} \backslash \{ \mathbf{v}_p, \bar{\mathbf{v}}_p  \}$ is a set of $n$ linearly independent vectors. If $\hat{V}_0$ is a set of linearly independent vectors, follow Step 6; otherwise skip Step 6 and follow Steps 7 through 9 to find $\mathbf{V}$ and $\mathbf{Z}$. 

\item [6)] Construct the matrix $\mathbf{V}$ from $\mathbf{V}_0$ by replacing the columns having $\mathbf{v}_p$ and $\bar{\mathbf{v}}_p$ with $\hat{\mathbf{v}}_p$ and $\bar{\hat{\mathbf{v}}}_p$ respectively. Similarly, construct the matrix $\mathbf{Z}$ from $\mathbf{Z}_0$ by replacing the corresponding columns of $\mathbf{Z}_0$ with $\mathbf{z}_p$ and $\bar{\mathbf{z}}_p$ respectively. Then jump to Step 10.

\item [7)] Find the largest-cardinality subset $V_1$ of $\hat{V}_0$ such that $\hat{\mathbf{v}}_p$, $\bar{\hat{\mathbf{v}}}_p \in V_1$ and $V_1$ is a self-conjugate set of linearly independent vectors.

\item [8)] This step is the same as the Step 8 of Sub-Algorithm 1. Thus, find $\hat{\mathbf{v}}_k$ in the column space of $\mathbf{N}_1(\lambda_k)$ for all $\mathbf{v}_k \in \hat{V}_0 \backslash V_1$ such that the set $\hat{V}= V_1 \cup \{\hat{\mathbf{v}}_k | \mathbf{v}_k \in \hat{V}_0 \backslash V_1 \}$ is a self conjugate set of $n$ linearly independent vectors, and find the corresponding $\mathbf{z}_k$. 

\item [9)] Construct $\mathbf{V}$ from $\mathbf{V}_0$ by replacing the columns containing $\mathbf{v}_p$, $\bar{\mathbf{v}}_p$  and all $\mathbf{v}_k\in \hat{V}_0 \backslash V_1$ with $\hat{\mathbf{v}}_p$, $\bar{\hat{\mathbf{v}}}_p$ and corresponding $\hat{\mathbf{v}}_k$ respectively. In the same manner, construct $\mathbf{Z}$ from $\mathbf{Z}_0$ by replacing the corresponding columns of $\mathbf{Z}_0$ with  $\mathbf{z}_p$, $\bar{\mathbf{z}}_p$ and all $\mathbf{z}_k$ obtained in Step 8 respectively.  

\item [10)] As before, compute the gain matrix $\mathbf{F}$ using (\ref{eq3a}).
\end{enumerate}

Now we formalize the outcome of Algorithm 2 in the following theorem:

\medskip

\noindent \textbf{Theorem 2:}  \textit{Consider the network synchronization model. Assume that: 1) the eigenvalues of $\mathbf{L}$ are distinct, 2) the number of actuation nodes is two more than the number of measurement nodes ($q=m+2$), and 3) the pair ($-\mathbf{L},\mathbf{B})$ is controllable. Then the gain matrix $\mathbf{F}$ of the state feedback controller obtained using Algorithm 2 blocks the observability of the model, specifically the pair $(\mathbf{C},-(\mathbf{L}+\mathbf{BF}))$ has an unobservable mode at $(-\lambda_p)$. Furthermore, all the open-loop eigenvalues and the open-loop eigenvectors in the set $V_0 \cap V_1$ are maintained in the closed-loop model.}

\medskip

\noindent \textbf{Proof:}

The proof is similar to that of Theorem 1, and thus is presented briefly to avoid repetition. We consider two cases separately for the proof: Case 1 where $\lambda_p$ is real and Case 2 where $\lambda_p$ is not real.

Recall that the steps of Sub-Algorithm 1 are followed for the Case 1. The key steps of Sub-Algorithm 1, the Step 7 and Step 8, differ from the those of Algorithm 1; hence, here we focus on showing that $V_1$ and $\hat{V}$ can be found respectively in Step 7 and Step 8 of Sub-Algorithm 1, verifying that the algorithm achieves a valid design. Since the open-loop eigenvectors are in conjugate pairs, it is immediate that in Step 7 a self-conjugate set $V_1$ can be found consisting of at least $(n-3)$ open-loop eigenvectors. Now we guarantee existence of $\hat{V}$ in Step 8 by using Proposition 1. To do so, consider $L_1$ to be the set of eigenvalues of $\mathbf{L}$ associated with the eigenvectors in $V_1$, and let $L_2= L_0 \backslash L_1$. Since $L_1, V_1$ and $L_2$ satisfy all the conditions of Proposition 1, therefore Proposition 1 can be applied which guarantees existence of $\hat{\mathbf{v}}_k$ in the column space of $\mathbf{N}_1(\lambda_k)$ for all $\lambda_k \in L_2$ such that $V_1 \cup \{\hat{\mathbf{v}}_k | \lambda_k \in L_2 \}$ is a self-conjugate set of $n$ linearly independent vectors. Hence, $\hat{V}$ in Step 8 can always be found. The remainder of the proof for this case is similar to that of Theorem 1.

Now consider the Case 2. Recall that the steps of Sub-Algorithm 2 are followed for this case. For this case we again will show the existence of $V_1$ and $\hat{V}$ in the Step 7 and Step 8 of Sub-Algorithm 2, respectively. If the vectors $\hat{\mathbf{v}}_p$ and $\bar{\hat{\mathbf{v}}}_p$ are linearly independent vectors, it is straightforward that the set $V_1$ consisting of at least $(n-6)$ open-loop eigenvectors can always be found in Step 7. Then, it remains to show that linearly independent vectors $\hat{\mathbf{v}}_p$ and $\bar{\hat{\mathbf{v}}}_p$ exist in Step 4 for any $\lambda_p$. Since $q=m+2$, there exist two linearly independent vectors in the null space of $\mathbf{N}_4(\lambda_p)$. As $\mathbf{N}_1(\lambda_p)$ is full-rank, therefore using suitable linear combination of the two linearly independent vectors in the null space of $\mathbf{N}_4(\lambda_p)$ a complex vector $\hat{\mathbf{v}}_p$ can always be obtained in Step 4 such that $\hat{\mathbf{v}}_p$ and $\bar{\hat{\mathbf{v}}}_p$ are linearly independent. Thus, $V_1$ exists in Step 7. Now the arguments in Case 1 can be directly applied to guarantee the existence of $\hat{V}$ in Step 8 of Sub-Algorithm 2. The remainder of the proof for this case is also similar to that of Theorem 1.

Finally note that $\mathbf{F}$ obtained in Algorithm 2 is a real valued matrix. This can be verified by standard manipulation of linear equations with conjugated coefficients (see \cite{jointeig} for further details). \hfill $\blacksquare$

\medskip

According to the theorem, the closed-loop model preserves the open-loop eigenvectors contained in the set $V_0 \cap V_1$. From the arguments in the proof it is clear that $V_0 \cap V_1$ contains at least $(n-6)$ open-loop eigenvectors (and at least $(n-3)$ open-loop eigenvectors when $\lambda_p$ is real). Therefore Algorithm 2 modifies only six open-loop eigenvectors at the worst. 

The matrices $\mathbf{V}$ and $\mathbf{Z}$ in Algorithm 2 are in general complex valued. Thus, calculation of $\mathbf{F}$ using (\ref{eq3a}) may not be computationally appealing as it involves inversion of complex valued matrix. However, an equivalent computation of $\mathbf{F}$ is possible, which involves manipulation of real valued matrices only and thus is computationally simpler. Here is the computation: $\mathbf{F} = \mathbf{Z}_{mod}~\mathbf{V}_{mod}^{-1}$ where $\mathbf{V}_{mod}$ and $\mathbf{Z}_{mod}$ are real matrices constructed from $\mathbf{V}$ and $\mathbf{Z}$ by replacing their complex conjugate columns with their real and imaginary parts, respectively. 

It should be noted that if our chosen $\lambda_p$ is real, we can use $q=m+1$ actuation nodes in Algorithm 2. However, the following example shows that Algorithm 2 may fail if $\lambda_p$ is not real and $q=m+1$ actuation nodes are used.  Therefore $q=m+2$ actuation nodes are required in general. 

{\em Example:} Consider the network model with parameters $\mathbf{L}= [0.9, -0.8, -0.1, 0;$ $-0.1, 0.9, -0.8, 0;$ $-0.8, -0.1, 1, -0.1;$ $0, 0, -0.2, 0.2]$, $\mathbf{B}=[1, 0, 0;$ $0, 1, 0;$ $0, 0, 0;$ $0, 0, 1]$ and $\mathbf{C}= [0, 0, 1, 0;$ $0, 0, 0, 1]$. Here, $m=2$ and $q=3$. For this network model, if we choose any of the complex eigenvalues of $\mathbf{L}$ as $\lambda_p$ in the Step 1, then $\mathbf{\hat{v}}_p$ obtained in the Step 4 will be real.\footnote{It can be shown that when each node in the network is either an actuation node or a measurement node, $\mathbf{\hat{v}}_p$ obtained in Step 4 is real.} Therefore, $V_1$ in Step 7 of Sub-Algorithm 2 would not exist, as $\bar{\hat{\mathbf{v}}}_p = \mathbf{\hat{v}}_p$. Hence, Algorithm 2 fails when $q=m+1$ actuation nodes are used. 

Next, we discuss the extension of our observablity-blocking controller design algorithm for the case when $\mathbf{L}$ has repeated eigenvalues. The extension is based on the results on eigenstructure assignment for repeated eigenvalues \cite{jointeig2,surgeig}. Below we briefly explain how this can be achieved by highlighting how the algorithms presented for the distinct eigenvalue case would change. 

\noindent \textbf{Brief discussion on the repeated eigenvalue case:} 

Repeated eigenvalues may be defective (have smaller geometric multiplicity than algebraic multiplicty), and hence may have Jordan chains which include generalized eigenvectors. In this case, the controller design algorithm must be modified to maintain the eigenvector and generalized eigenvector sequences of the Jordan chains. Thus, when any eigenvector/generalized eigenvector in a Jordan chain is modified in Step 4 or Step 8, all the following generalized eigenvectors in that Jordan chain must also be modified accordingly in that step. The modified/target closed-loop generalized eigenvectors and associated $\mathbf{z}$ vectors can be obtained through condition 3 of \cite{jointeig2}, which characterizes the assignability of generalized eigenvectors via state feedback. As the sequences in the Jordan chains are maintained, the technique for surgical eigenstructure assignment in \cite{surgeig} can again be applied to obtain target closed-loop eigenvectors/generalized eigenvectors in Step 8. Finally the gain matrix of the observability-blocking controller can be obtained using (\ref{eq3a}). It should be noted that, like Algorithm 2, conjugate open-loop eigenvectors/generalized eigenvectors must also be modified accordingly when $\mathbf{L}$ has complex eigenvalues.

The controller designed according to the above discussion blocks observability and is guaranteed to maintain all the open-loop eigenvalues in the closed-loop model. However, in general, it is difficult to characterize how much of the open-loop eigenvectors/generalized eigenvectors will be maintained. For the sake of completeness, we formalize the discussion in the following theorem. The proof is omitted (it is similar to the previous proofs but require the results in \cite{jointeig2,surgeig}).

\medskip

\noindent \textbf{Theorem 3:}  \textit{Consider the network synchronization model. Assume that: 1) the number of actuation nodes is two more than the number of measurement nodes ($q=m+2$), and 2) the pair ($-\mathbf{L},\mathbf{B})$ is controllable. Then a state feedback gain matrix $\mathbf{F}$ can be designed so that observability is blocked \big(i.e. the pair $(\mathbf{C},-(\mathbf{L}+\mathbf{BF}))$ is  unobservable, where any chosen open-loop mode $(-\lambda_p)$ can be made unobservable\big) and the open-loop eigenvalues are maintained in the closed-loop model.}

\medskip

{\em Remark:}The algorithms presented so far in this section are based on an application of the eigenstructure assignment techniques in \cite{jointeig,jointeig2,surgeig}.  Relative to\cite{jointeig,jointeig2,surgeig}, the contribution and effort in this study lie in designing the target eigenvectors to block observability and preserve most of the eigenstructure, while meeting the criteria for assignability given in \cite{jointeig,jointeig2,surgeig}. From another viewpoint,  \cite{jointeig,jointeig2,surgeig} give implicit conditions that can be checked for assignability; our effort here shows how the target eigenstructure set can be designed to meet these conditions.

Several further remarks on the observability-blocking controller design algorithm are worthwhile:

\begin{enumerate}

\item The presented design method is general, in the sense that it does not depend on the state matrix having a Laplacian form, nor on the specific graph topolgy.  In Section IV.B and IV.C, the design will be specialized for sparser and regional feedback controls, by exploiting the graph topology and Laplacian form of the network dynamics.

\item If the network graph is undirected or is a tree, observability can be blocked using $q=m+1$ actuation nodes, as the Laplacians have only real eigenvalues in these cases. 

\item The controller designed using the algorithm maintains almost the entire open-loop eigenstructure. Such a structured modification of the eigenstructure allows for sequential design of feedback controllers for additional goals beyond unobservability. As an example, controllers could be first designed for a performance goal (e.g. eigenvalue placement), and then a second control loop could be applied to block observability while maintaining the previously placed eigenvalues. In addition, maintenance of eigenstructure may be appealing from a privacy standpoint, in the sense that other stakeholders may not be able to distinguish the presence of the observability-blocking controller.

\item The design method allows the control designer to choose which mode is made unobservable.  The mode can be chosen, for example, based on the achieved unobservable subspace or the extent to which the open-loop dynamics is modified.

\item Multiple modes of the system dynamics can be made unobservable using the design method, by modifying Step 4 so that all the corresponding eigenvectors have the necessary zero patterns. Alternatively, sequential design can also be used to enforce unobservability on multiple modes. At most $(n-m)$ modes can be made unobservable.

\item The presented design method can be extended for uncontrollable systems. The main idea is that all the target eignvectors/generalized eigenvectors must be chosen to maintain the open-loop left eigenvectors associated with all uncontrollable eigenvalues.

\item We can also design a controller to enable observability of a mode $(-\lambda_p)$ for which the open-loop model is unobservable. In that case, $\mathbf{h}_p$ must be selected in Step 3 in such a way that $\mathbf{N}_4(\lambda_p)~ \mathbf{h}_p \neq \mathbf{0}$. By using $(n-m+1)$ actuation channels, we can guarantee the existence of such an $\mathbf{h}_p$, however many fewer controllers often suffice. Once $\mathbf{h}_p$ is selected, the remainder of the design is similar to that for the observability-blocking control.

\end{enumerate}

\subsection{Sparser Observability-Blocking Using Network Graph Cutsets}

In the previous section, we showed that observability can be blocked at a set of measurement nodes, by applying state feedback at two more nodes than there are measurement nodes. By exploiting the topological structure of the Laplacian matrix, it turns out that we can often block observability using state feedback at a smaller set of network nodes. The genesis of this sparser design is that blocking observability on the nodes associated to a vertex-cutset of the network graph using actuation in one partition can serve to block observability at all nodes associated with the other partition. We formalize this notion first, and then discuss how the idea can be used to obtain sparser observability-blocking designs.

\begin{figure}[thpb] 
\centering
\includegraphics[width=8cm,height=4.5cm]{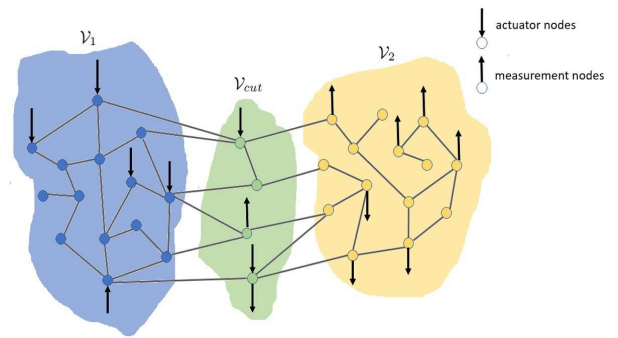}
\caption{Network graph $\mathcal{G}$ with the vertex-sets considered in the sparser design.} \label{fig1}
\end{figure}

To formalize the notion, it is helpful to explicitly define two synchronization network models which have different measurement paradigms. Specifically, we refer to the network synchronization model defined in Section II as the {\em base synchronization network model}.
We also consider a second model which has the same state dynamics and actuation nodes as the base model (hence the state matrix is $\mathbf{L}$ and the input matrix is $\mathbf{B}$), however the measurement model is different. Specifically, to define the measurement model, we consider a vertex-cutset of the network graph that separates actuation vertices and  measurement vertices in the base model, as illustrated in Fig. \ref{fig1}. In the figure, the cutset $\mathcal{V}_{cut}$ partitions the graph into two vertex-sets $\mathcal{V}_1$ and $\mathcal{V}_2$ such that $\mathcal{V}_1$ does not include any measurement vertex, $\mathcal{V}_2$ does not include any actuation vertex and there is no edge between any vertex in $\mathcal{V}_1$ and any vertex in $\mathcal{V}_2$. Note that $\mathcal{V}_1, \mathcal{V}_{cut}$ and $\mathcal{V}_2$ are mutually exclusive and $\mathcal{V}_1 \cup \mathcal{V}_{cut} \cup \mathcal{V}_2 = \mathcal{V}$. We stress that the cutset may include the actuation vertices or measurement vertices of the base model as shown in the figure, and hence a cutset always exists such that $|\mathcal{V}_{cut}| \leq m$ where $|\mathcal{V}_{cut}|$ represents the cardinality of $\mathcal{V}_{cut}$. (In the extreme case, the vertex-cutset may include all the base model's measurement vertices). In the second model, new measurement vertices are defined as the vertices in the cutset $\mathcal{V}_{cut}$ where $|\mathcal{V}_{cut}| \leq m$, and the corresponding output matrix is defined as $\mathbf{\hat{C}}$ to differentiate it from the base model; we refer to this second model as the {\em cutset-measurement synchronization network model}. 

Without loss of generality, for our convenience we renumber the vertices and associated nodes as follows. We assign indices to the vertices in $\mathcal{V}_1$ first, then vertices in $\mathcal{V}_{cut}$ and lastly vertices in $\mathcal{V}_2$. Therefore, the Laplacian $\mathbf{L}$ can now be partitioned into blocks as
\begin{eqnarray} \label{block_L}
\mathbf{L}=
  \begin{bmatrix}
    \mathbf{L}_{\mathcal{V}_1 \mathcal{V}_1} & \mathbf{L}_{\mathcal{V}_1\mathcal{V}_{cut}} & \mathbf{0} \\
    \mathbf{L}_{\mathcal{V}_{cut}\mathcal{V}_1} & \mathbf{L}_{\mathcal{V}_{cut}\mathcal{V}_{cut}} & \mathbf{L}_{\mathcal{V}_{cut}\mathcal{V}_2}\\
    \mathbf{0} & \mathbf{L}_{\mathcal{V}_2\mathcal{V}_{cut}} & \mathbf{L}_{\mathcal{V}_2\mathcal{V}_2}
  \end{bmatrix}
\end{eqnarray}
where e.g. $\mathbf{L}_{\mathcal{V}_1\mathcal{V}_{cut}}$ refers to the block of $\mathbf{L}$ whose rows and columns correspond to the vertices of $\mathcal{V}_1$ and  $\mathcal{V}_{cut}$ respectively (other blocks are referenced similarly). 

The following lemma gives a condition under which observability-blocking in the cutset-measurement synchronization network model implies observability-blocking in the base synchronization network model. This lemma will be used in tandem with our design algorithm presented in Section IV.A to obtain a sparser observability-blocking controller design.

\medskip

\noindent \textbf{Lemma 1:} \textit{Consider a base synchronization network model and associated cutset-measurement synchronization network model. Suppose the gain matrix $\mathbf{F}$ of the state feedback controller is designed to block observability in the cutset-measurement synchronization network model, in such a way the pair $(\mathbf{\hat{C}},-(\mathbf{L}+\mathbf{BF}))$ has an unobservable mode at $(-\lambda_p)$ where $\lambda_p$ is not an eigenvalue of the block $\mathbf{L}_{\mathcal{V}_2\mathcal{V}_2}$. Then this feedback controller also serves to block observability in the base synchronization network model, specifically the pair $(\mathbf{C},-(\mathbf{L}+\mathbf{BF}))$ has an unobservable mode at $(-\lambda_p)$.}

\medskip

\noindent \textbf{Proof:}

Consider a synchronization network defined on graph $\mathcal{G}=(\mathcal{V},\mathcal{E}: \mathcal{W})$ and also consider the vertex-sets $\mathcal{V}_1, \mathcal{V}_{cut}$ and $\mathcal{V}_2$ (see Fig. \ref{fig1}), as defined above. Note the proof is immediate for the trivial case where $\mathcal{V}_2= \emptyset$, because in that case $\mathbf{\hat{C}}= \mathbf{C}$. Now consider that $\mathcal{V}_2 \neq \emptyset$. Suppose $\mathbf{\hat{v}}_p$ is the eigenvector of $\mathbf{L}+\mathbf{BF}$ for the eigenvalue at $\lambda_p$. Since ($-\lambda_p$) is the unobserable mode of closed-loop cutset-measurement synchronization network model, all the entries of $\mathbf{\hat{v}}_p$ corresponding to the vertices in the cutset are zero. Hence we can write $\mathbf{\hat{v}}_p=[\mathbf{\hat{v}}^T_{p_{\mathcal{V}_1}} ~ \mathbf{0}^T ~ \mathbf{\hat{v}}^T_{p_{\mathcal{V}_2}} ]^T$. Now we expand the eigenvector equation of $\lambda_p$ to characterize the eigenvector's zero pattern. Specifically, using the block expression of the Laplacian given in (\ref{block_L}), we can re-write $\mathbf{(L+BF)}\mathbf{\hat{v}}_p = \lambda_p \mathbf{\hat{v}}_p$ as
 \begin{eqnarray} \label{eq6}
\resizebox{.9\hsize}{!}
{$\begin{bmatrix}
    \mathbf{L}_{\mathcal{V}_1 \mathcal{V}_1}+ \mathbf{F}_{\mathcal{V}_1 \mathcal{V}_1} & \mathbf{L}_{\mathcal{V}_1\mathcal{V}_{cut}}+\mathbf{F}_{\mathcal{V}_1\mathcal{V}_{cut}} & \mathbf{F}_{\mathcal{V}_1\mathcal{V}_2} \\
    \mathbf{L}_{\mathcal{V}_{cut}\mathcal{V}_1} + \mathbf{F}_{\mathcal{V}_{cut} \mathcal{V}_1} & \mathbf{L}_{\mathcal{V}_{cut}\mathcal{V}_{cut}} +\mathbf{F}_{\mathcal{V}_{cut}\mathcal{V}_{cut}}& \mathbf{L}_{\mathcal{V}_{cut}\mathcal{V}_2}+\mathbf{F}_{\mathcal{V}_{cut}\mathcal{V}_2}\\
    \mathbf{0} & \mathbf{L}_{\mathcal{V}_2\mathcal{V}_{cut}} & \mathbf{L}_{\mathcal{V}_2\mathcal{V}_2}
  \end{bmatrix}
   \begin{bmatrix}
    \mathbf{\hat{v}}_{p_{\mathcal{V}_1}}  \\
   \mathbf{0} \\
   \mathbf{\hat{v}}_{p_{\mathcal{V}_2}}
  \end{bmatrix}
 = \lambda_p
 \begin{bmatrix}
    \mathbf{\hat{v}}_{p_{\mathcal{V}_1}}  \\
   \mathbf{0} \\
   \mathbf{\hat{v}}_{p_{\mathcal{V}_2}}
  \end{bmatrix}$}
  \end{eqnarray}
Here the feedback $\mathbf{F}$ is partitioned according to the defined vertex-sets, e.g. $\mathbf{F}_{\mathcal{V}_1 \mathcal{V}_2}$ refers to the feedback applied to  $\mathcal{V}_1$ using measurements from $\mathcal{V}_2$ through state feedback. From the bottom block of (\ref{eq6}) we obtain $\mathbf{L}_{\mathcal{V}_2\mathcal{V}_2}~ \mathbf{\hat{v}}_{p_{\mathcal{V}_2}}= \lambda_{p} \mathbf{\hat{v}}_{p_{\mathcal{V}_2}}$. This implies $\mathbf{\hat{v}}_{p_{\mathcal{V}_2}}= \mathbf{0}$ since $\lambda_p$ is not an eigenvalue of the block $\mathbf{L}_{\mathcal{V}_2\mathcal{V}_2}$ according to our assumption. Now note that $\mathcal{V}_2 \cup \mathcal{V}_{cut}$ contains all the measurement vertices and the entries of $\mathbf{\hat{v}}_p$ corresponding to the vertices in $\mathcal{V}_2 \cup \mathcal{V}_{cut}$ are zero. Hence the proof is complete. \hfill $\blacksquare$

\medskip

Lemma 1 serves as a basis for an algorithm for sparser observability-blocking control.  The main idea is to use the algorithms from Section IV.A to design an observability-blocking controller for the cutset-measurement synchronization network model (which may have fewer measurement nodes than the base model), whereupon Lemma 1 can be leveraged to guarantee that observability is blocked in the original model. Specifically, this can be done as follows. First, a small-cardinality vertex-cutset $\mathcal{V}_{cut}$ separating the measurement vertices from the actuation vertices in the network graph is chosen, and the associated cutset-measurement synchronization network model is formed. Then, in accordance with the algorithms in Section IV.A, an eigenvalue $\lambda_p$ of $\mathbf{L}$ is chosen, whose eigenvector will be modified to block observability; however, an additional technical criterion is imposed on the selection, that $\lambda_p$ is not an eigenvalue of the block $\mathbf{L}_{\mathcal{V}_2\mathcal{V}_2}$. Thereupon, the algorithms presented in Section IV.A can be applied to find the controller that makes the mode at ($-\lambda_p$) unobservable in the cutset-measurement synchronization network model using $|\mathcal{V}_{cut}|+2$ actuation nodes. From Lemma 1, it is immediate that the same controller blocks observability in the base synchronization network model. In this way the desired sparser observability-blocking controller is found that requires only $|\mathcal{V}_{cut}|+2$ actuation nodes. Further, this controller maintains all of the open-loop eigenvalues in the closed-loop, according to Theorem 3. 

The above design requires $\mathbf{L}$ to have at least one eigenvalue which is not an eigenvalue of the block $\mathbf{L}_{\mathcal{V}_2\mathcal{V}_2}$. We claim that there is such an eigenvalue, when the network graph is strongly connected. In particular, $\mathbf{L}$ has an eigenvalue at zero, but under the strong connectivity condition,  $\mathbf{L}_{\mathcal{V}_2\mathcal{V}_2}$ cannot have an eigenvalue at zero. This is because $\mathbf{L}_{\mathcal{V}_2\mathcal{V}_2}$ is a grounded Laplacian where every vertex in  $\mathcal{V}_2$ can be reached through a directed path from some vertex in $\mathcal{V}_1 \cup \mathcal{V}_{cut}$, and hence all the eigenvalues of $\mathbf{L}_{\mathcal{V}_2\mathcal{V}_2}$ are in the open right half plane (see Theorem 1 of \cite{gl_1}).

We summarize this result in the following theorem. The proof is omitted as it is immediate from the above discussion.

\medskip
 
\noindent \textbf{Theorem 4:} \textit{Consider the  synchronization network model, and say that a separating vertex-cutset $V_{cut}$ has been found.  Assume that: 1) the number of actuation nodes is two more than the cardinality of the vertex-cutset $\mathcal{V}_{cut}$ ($q=|\mathcal{V}_{cut}|+2$), and 2) the pair ($-\mathbf{L},\mathbf{B})$ is controllable. Then the gain matrix $\mathbf{F}$ of state feedback controller can be designed, according to the process described above, to block observability  (i.e. to make the pair $(\mathbf{C},-(\mathbf{L}+\mathbf{BF}))$ unobservable). Furthermore, for this design, all the open-loop eigenvalues are maintained in the closed-loop system.}

\medskip

{\em Remark:} When all the eigenvalues of $\mathbf{L}$ are real, we can use one fewer actuation node  ($q=|\mathcal{V}_{cut}|+1$ actuation nodes) in our design.

The sparser observability-blocking controller design formalized in Theorem 4 is especially useful when the cardinality of the cutset is much less than the number of measurement nodes $m$ (i.e. $|\mathcal{V}_{cut}| << m$). Graphs for many real-world networks are sparse, and  have small-cardinality vertex-cutsets.  For instance, the power grid is known to have a sparse structure with regions in the grid connected by few long transmission lines, while backbones for traffic and communication infrastructures are planar and often quite sparse. Networks in nature with scale-free structures are also essentially sparse \cite{sprs}. In such networks, Theorem 4 indicates that observability can be blocked, and hence security against adversaries can be enhanced, using only a few actuators compared to the number of measurement nodes. As a limiting case, if the network graph is tree, we can block observability in a connected subnetwork using only two actuators, assuming that the model is controllable. 

\subsection{Observability-Blocking Using Regional State feedback Controllers}

Thus far, we have considered the design of state feedback controllers to block observability. However, because observability-blocking control problems tend to arise in networks with independent or adversarial control authorities, these authorities sometimes may not have access to the full network state. Here, we consider the alternative that control authorities only can access states in a region of the network. For this case, we demonstrate that observability can be blocked at measurement nodes even outside the region, by blocking observability at nodes associated with a vertex-cutset at the boundary of the accessible region. A general design of this sort, which however does not necessarily maintain stability, is demonstrated in Theorem 5. Then stability-preserving design is obtained by applying a time-scale separating control (Theorem 6).

Formally, we assume that there is a subset of nodes in the network whose states can be accessed and used in feedback by the control authorities; this subset is assumed to include all actuation nodes. We refer to these nodes as {\em accessible} nodes (respectively accessible vertices in the graph), and the remaining as {\em inaccessible} nodes (respectively vertices).

\begin{figure}[thpb] 
\centering
\includegraphics[width=8cm,height=4.5cm]{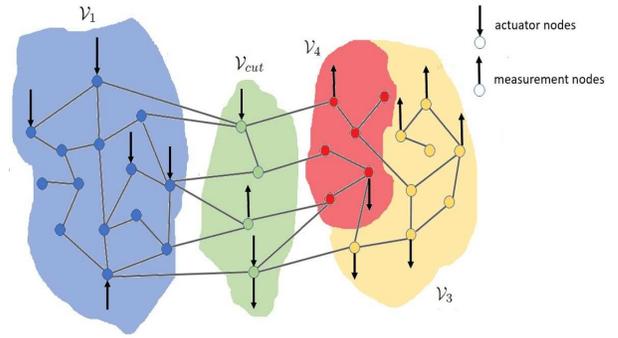}
\caption{Network graph $\mathcal{G}$ with the vertex-sets considered in the regional state feedback controller design.} \label{fig2}
\end{figure}

To design observability-blocking controllers, it is again helpful to consider two synchronization network models. The first model is the base synchronization network model defined in Section IV.B. The second model resembles the measurement-cutset synchronization network model, but only encompasses the subnetwork of accessible nodes, and also defines a separating cutset based on accessibility.  

To develop the second model, we consider a vertex-cutset $\mathcal{V}_{cut}$ of the network graph, which creates the following partitions (see Fig. \ref{fig2}): 1) a partition $\mathcal{V}_1$ with only accessible vertices which does not contain any measurement vertices; and 2) a partition $\mathcal{V}_2$ with other vertices which includes all the inaccessible vertices (and may include accessible vertices) and does not contain any actuation vertices. Here, the cutset can contain actuation and measurement vertices, but all vertices in the cutset must be accessible. It is easy to check such a cutset can always be found. Next, we distinguish the accessible and inaccessible vertices in $\mathcal{V}_2$ as $\mathcal{V}_3$ and $\mathcal{V}_4$ respectively. The model is defined on the accessible vertices i.e. vertex-set $\mathcal{V}_1 \cup \mathcal{V}_{cut} \cup \mathcal{V}_3$. Formally, the network graph of the second model is the subgraph of $\mathcal{G}$ induced by the vertex-set $\mathcal{V}_1 \cup \mathcal{V}_{cut} \cup \mathcal{V}_3$ (see \cite{gt_1} for background on induced subgraphs). We denote this subgraph and associated Laplacian matrix as $\mathcal{\tilde{G}}$ and $\mathbf{\tilde{L}}$ respectively. This model is defined as having the same actuation vertices as the base model.  However, the measurement vertices are defined differently, as the vertices in  $\mathcal{V}_{cut}$. We denote the input and output matrix as $\mathbf{\tilde{B}}$ and $\mathbf{\tilde{C}}$ respectively, and formally refer to this model as {\em accessible-region synchronization network model}. For convenience, we call the network defined on the subgraph of $\mathcal{G}$ induced by the remaining inaccessible vertex-set $\mathcal{V}_4$ as the {\em inaccessible-region synchronization network} and denote the associated Laplacian matrix as $\mathbf{\tilde{L}}_{rem}$.

Without loss of generality, we renumber the vertices/nodes as follows. We assign indices to the vertices in $\mathcal{V}_1$ first, then vertices in $\mathcal{V}_{cut}$, then vertices in $\mathcal{V}_3$, and lastly vertices in $\mathcal{V}_4$. Based on this indexing, the Laplacian $\mathbf{L}$ can be partitioned into blocks as
\begin{eqnarray} \label{Lap_block1} 
\mathbf{L}=
  \begin{bmatrix}
    \mathbf{L}_{\mathcal{V}_1 \mathcal{V}_1} & \mathbf{L}_{\mathcal{V}_1\mathcal{V}_{cut}} & \mathbf{0} & \mathbf{0}\\
    \mathbf{L}_{\mathcal{V}_{cut}\mathcal{V}_1} & \mathbf{L}_{\mathcal{V}_{cut}\mathcal{V}_{cut}} & \mathbf{L}_{\mathcal{V}_{cut}\mathcal{V}_3} & \mathbf{L}_{\mathcal{V}_{cut}\mathcal{V}_4}\\
    \mathbf{0} & \mathbf{L}_{\mathcal{V}_3\mathcal{V}_{cut}} & \mathbf{L}_{\mathcal{V}_3\mathcal{V}_3} & \mathbf{L}_{\mathcal{V}_3\mathcal{V}_4}\\
    \mathbf{0} & \mathbf{L}_{\mathcal{V}_4\mathcal{V}_{cut}} & \mathbf{L}_{\mathcal{V}_4\mathcal{V}_3} & \mathbf{L}_{\mathcal{V}_4\mathcal{V}_4}\\
  \end{bmatrix}
\end{eqnarray}
where e.g. $\mathbf{L}_{\mathcal{V}_1\mathcal{V}_{cut}}$ refers to the block of $\mathbf{L}$ whose rows and columns correspond to the vertices of $\mathcal{V}_1$ and $\mathcal{V}_{cut}$ respectively (and other blocks are referenced similarly). Similarly, the Laplacian $\mathbf{\tilde{L}}$ of the accessible-region synchronization network model can also written as
\begin{eqnarray}
\resizebox{1\hsize}{!}
{$
\mathbf{\tilde{L}}=
  \begin{bmatrix}
    \mathbf{\tilde{L}}_{\mathcal{V}_1 \mathcal{V}_1} & \mathbf{\tilde{L}}_{\mathcal{V}_1\mathcal{V}_{cut}} & \mathbf{0} \\
    \mathbf{\tilde{L}}_{\mathcal{V}_{cut}\mathcal{V}_1} & \mathbf{\tilde{L}}_{\mathcal{V}_{cut}\mathcal{V}_{cut}} & \mathbf{\tilde{L}}_{\mathcal{V}_{cut}\mathcal{V}_3} \\
    \mathbf{0} & \mathbf{\tilde{L}}_{\mathcal{V}_3\mathcal{V}_{cut}} & \mathbf{\tilde{\tilde{L}}}_{\mathcal{V}_3\mathcal{V}_3}
  \end{bmatrix}= 
  \begin{bmatrix}
    \mathbf{L}_{\mathcal{V}_1 \mathcal{V}_1} & \mathbf{L}_{\mathcal{V}_1\mathcal{V}_{cut}} & \mathbf{0} \\
    \mathbf{L}_{\mathcal{V}_{cut}\mathcal{V}_1} & \mathbf{\tilde{L}}_{\mathcal{V}_{cut}\mathcal{V}_{cut}} & \mathbf{L}_{\mathcal{V}_{cut}\mathcal{V}_3} \\
    \mathbf{0} & \mathbf{L}_{\mathcal{V}_3\mathcal{V}_{cut}} & \mathbf{\tilde{L}}_{\mathcal{V}_3\mathcal{V}_3}
  \end{bmatrix}\nonumber $}
\end{eqnarray}
\vspace{-0.5cm}
\begin{eqnarray} \label{Lap_block2} 
\end{eqnarray}
where e.g. $\mathbf{\tilde{L}}_{\mathcal{V}_1\mathcal{V}_{cut}}$ refers to the block of $\mathbf{\tilde{L}}$ whose rows and columns correspond to the vertices of $\mathcal{V}_1$ and $\mathcal{V}_{cut}$ respectively (and the other blocks are referenced similarly). 

The following lemma gives a condition under which an observability-blocking controller designed for the accessible-region synchronization network model yields an observability-blocking controller for the base synchronization network model. This result is then used in tandem with the observability-blocking controller design presented in Section IV.A to obtain a general deign for the regional state feedback case.

\medskip

\noindent \textbf{Lemma 2:} \textit{Consider a base synchronization network model and associated accessible-region synchronization network model. Suppose the gain matrix $\mathbf{\tilde{F}}$ of the state feedback controller is designed to block observability in the accessible-region synchronization network model, in such a way that the pair $(\mathbf{\tilde{C}},-(\mathbf{\tilde{L}}+\mathbf{\tilde{B}\tilde{F}}))$ has an unobservable mode at $(-\tilde{\lambda}_p)$ where $\tilde{\lambda}_p$ is not an eigenvalue of the block $\mathbf{\tilde{L}}_{\mathcal{V}_3\mathcal{V}_3}$. Then the gain matrix $\mathbf{F}$ obtained from $\mathbf{\tilde{F}}$ via appropriate zero padding \big(i.e. $\mathbf{F}= [\mathbf{\tilde{F}}~~ \mathbf{0}]$\big) blocks observability in the base synchronization network model, specifically the pair $(\mathbf{C},-(\mathbf{L}+\mathbf{BF}))$ has an unobservable mode at $(-\tilde{\lambda}_p)$.}

\medskip

\noindent \textbf{Proof:}

Consider a synchronization network defined on graph $\mathcal{G}=(\mathcal{V},\mathcal{E}: \mathcal{W})$ and also consider the vertex-sets $\mathcal{V}_1$, $\mathcal{V}_2$, $\mathcal{V}_{cut}$, $\mathcal{V}_3$ and $\mathcal{V}_4$ (see Fig. \ref{fig2}) as defined above. When $\mathcal{V}_4= \emptyset$, the proof is immediate from Lemma 1. 

Now consider the case that $\mathcal{V}_4 \neq \emptyset$ (i.e. the network has inaccessible nodes). Assume that $\mathbf{\tilde{v}}_p$ is the eigenvector of $\mathbf{\tilde{L}}+\mathbf{\tilde{B}\tilde{F}}$ for the eigenvalue at $\tilde{\lambda}_p$. Now we will apply Lemma 1 to characterize the zero pattern of $\mathbf{\tilde{v}}_p$. For that, first consider an interim model on the accessible region whose state dynamics and actuation vertices are same as the accessible-region synchronization network model but measurement vertices are different, which are $\mathcal{V}_{cut} \cup \mathcal{V}_3$. Note, $\mathcal{V}_{cut}$ serves as a cutset between actuation and measurement vertices of this interim model. By assumption, $\tilde{\lambda}_p$ is not an eigenvalue of the block $\mathbf{\tilde{L}}_{\mathcal{V}_3\mathcal{V}_3}$ and the gain matrix $\mathbf{\tilde{F}}$ is designed to block observability in the accessible-region synchronization network model so that the mode $(-\tilde{\lambda}_p)$ is unobservable. Hence, Lemma 1 can be applied here by comparing the accessible-region synchronization network model and the interim model respectively to the cutset-measurement synchronization network model and the base model of Lemma 1. Thereby per Lemma 1, $\mathbf{\tilde{F}}$ also serves to block observability in the interim model so that the mode $(-\tilde{\lambda}_p)$ is unobservable. As a consequence, all the entries of $\mathbf{\tilde{v}}_p$ corresponding to $\mathcal{V}_{cut} \cup \mathcal{V}_3$, the measurement vertices of the interim model, are zero. This implies $\mathbf{\tilde{v}}_p =[\mathbf{\tilde{v}}^T_{p_{\mathcal{V}_1}} ~ \mathbf{0}^T ~ \mathbf{0}^T]^T$ where $\mathbf{\tilde{v}}_{p_{\mathcal{V}_1}} \neq \mathbf{0}$. Now we can write the eigenvector equation for the eigenvalue $\tilde{\lambda}_p$ of the accessible-region synchronization network model as

\begin{eqnarray} \label{eq7}
\resizebox{0.9\hsize}{!}
{$\begin{bmatrix}
    \mathbf{L}_{\mathcal{V}_1 \mathcal{V}_1}+ \mathbf{\tilde{F}}_{\mathcal{V}_1 \mathcal{V}_1} & \mathbf{L}_{\mathcal{V}_1\mathcal{V}_{cut}}+\mathbf{\tilde{F}}_{\mathcal{V}_1\mathcal{V}_{cut}} & \mathbf{\tilde{F}}_{\mathcal{V}_1\mathcal{V}_3} \\
    \mathbf{L}_{\mathcal{V}_{cut}\mathcal{V}_1}+ \mathbf{\tilde{F}}_{\mathcal{V}_{cut}\mathcal{V}_1} & \mathbf{\tilde{L}}_{\mathcal{V}_{cut}\mathcal{V}_{cut}} + \mathbf{\tilde{F}}_{\mathcal{V}_{cut}\mathcal{V}_{cut}}& \mathbf{L}_{\mathcal{V}_{cut}\mathcal{V}_3}+ \mathbf{\tilde{F}}_{\mathcal{V}_{cut}\mathcal{V}_3}\\
    \mathbf{0} & \mathbf{L}_{\mathcal{V}_3\mathcal{V}_{cut}} & \mathbf{\tilde{L}}_{\mathcal{V}_3\mathcal{V}_3}
  \end{bmatrix}
   \begin{bmatrix}
    \mathbf{\tilde{v}}_{p_{\mathcal{V}_1}}  \\
   \mathbf{0} \\
   \mathbf{0}
  \end{bmatrix}
 = \tilde{\lambda}_p
 \begin{bmatrix}
    \mathbf{\tilde{v}}_{p_{\mathcal{V}_1}}  \\
   \mathbf{0} \\
  \mathbf{0}
  \end{bmatrix}$} .
  \end{eqnarray}
  
Here, the feedback $\mathbf{\tilde{F}}$ is partitioned according to the defined vertex-sets, e.g. $\mathbf{\tilde{F}}_{\mathcal{V}_1 \mathcal{V}_{cut}}$ refers to the feedback applied to $\mathcal{V}_1$ using measurements from $\mathcal{V}_{cut}$ through state feedback. Also note that $\mathbf{B}= [\mathbf{\tilde{B}}^T~~ \mathbf{0}^T]^T$. Let us consider $\mathbf{F}= [\mathbf{\tilde{F}}~ \mathbf{0}]$ and $\mathbf{v}_p=[\mathbf{\tilde{v}}_p^T~~ \mathbf{0}^T]^T$. Then using (\ref{Lap_block1}), (\ref{Lap_block2}) and (\ref{eq7}) we can write $\mathbf{(L+BF)}\mathbf{v}_p$ as
\begin{eqnarray} \label{eq8}
\resizebox{1\hsize}{!}
{$
\begin{bmatrix}
     \mathbf{L}_{\mathcal{V}_1 \mathcal{V}_1}+ \mathbf{\tilde{F}}_{\mathcal{V}_1 \mathcal{V}_1} & \mathbf{L}_{\mathcal{V}_1\mathcal{V}_{cut}}+\mathbf{\tilde{F}}_{\mathcal{V}_1\mathcal{V}_{cut}} & \mathbf{\tilde{F}}_{\mathcal{V}_1\mathcal{V}_3} & \mathbf{0} \\
    \mathbf{L}_{\mathcal{V}_{cut}\mathcal{V}_1}+ \mathbf{\tilde{F}}_{\mathcal{V}_{cut}\mathcal{V}_1} & \mathbf{L}_{\mathcal{V}_{cut}\mathcal{V}_{cut}}+ \mathbf{\tilde{F}}_{\mathcal{V}_{cut}\mathcal{V}_{cut}} & \mathbf{L}_{\mathcal{V}_{cut}\mathcal{V}_3} + \mathbf{\tilde{F}}_{\mathcal{V}_{cut}\mathcal{V}_3}& \mathbf{L}_{\mathcal{V}_{cut}\mathcal{V}_4}\\
    \mathbf{0} & \mathbf{L}_{\mathcal{V}_3\mathcal{V}_{cut}} & \mathbf{L}_{\mathcal{V}_3\mathcal{V}_3} & \mathbf{L}_{\mathcal{V}_3\mathcal{V}_4}\\
    \mathbf{0} & \mathbf{L}_{\mathcal{V}_4\mathcal{V}_{cut}} & \mathbf{L}_{\mathcal{V}_4\mathcal{V}_3} & \mathbf{L}_{\mathcal{V}_4\mathcal{V}_4}\\
  \end{bmatrix}
   \begin{bmatrix}
    \mathbf{\tilde{v}}_{p_{\mathcal{V}_1}}\\ 
   \mathbf{0} \\
   \mathbf{0} \\
   \mathbf{0}
  \end{bmatrix}
 = \tilde{\lambda}_p \mathbf{v}_p. $} \nonumber 
  \end{eqnarray}
\vspace{-0.5cm}
\begin{eqnarray} 
\end{eqnarray}
Therefore, $\tilde{\lambda}_p$ and $\mathbf{v}_p$ are the eigenvalue and associated eigenvector of $\mathbf{(L+BF)}$. Since all the entries of $\mathbf{v}_p$ corresponding to $\mathcal{V}_{cut} \cup \mathcal{V}_2$ (which includes all the measurement vertices) are zero, $(-\tilde{\lambda}_p)$ is an unobservable mode of the closed-loop base synchronization network model. Hence $\mathbf{F}= [\mathbf{\tilde{F}}~ \mathbf{0}]$ makes the base synchronization network model unobservable. \hfill $\blacksquare$

\medskip

Lemma 2 is the basis of our algorithm for designing regional state feedback observability-blocking controller. The main idea is to use the algorithm presented in Section IV.A to design $\mathbf{\tilde{F}}$ of the observability-blocking controller for the accessible-region synchronization network model, and then employ Lemma 2 to guarantee that $\mathbf{F}= [\mathbf{\tilde{F}}~~ \mathbf{0}]$ blocks observability in the original model. This controller serves as a regional state feedback controller since the entries of $\mathbf{F}$ associated to inaccessible nodes are zero and therefore only the states of the accessible nodes are used by this controller. Systematically, this controller can be obtained as follows. First, a vertex-cutset $\mathcal{V}_{cut}$ is chosen as defined above and the associated accessible-region synchronization network model is formed. Then, in accordance with the algorithms in Section IV.A, an eigenvalue $\tilde{\lambda}_p$ of $\mathbf{\tilde{L}}$ is chosen, whose eigenvector will be modified to block observability in the accessible-region synchronization network model; however, similar to the sparser observability-blocking controller design, an additional technical criterion is imposed on the selection, that $\tilde{\lambda}_p$ is not an eigenvalue of the block $\mathbf{\tilde{L}}_{\mathcal{V}_3\mathcal{V}_3}$. Next, the observability-blocking controller algorithm in Section IV.A is applied to find $\mathbf{\tilde{F}}$ that makes the mode at $(-\tilde{\lambda}_p)$ unobservable in the accessible-region synchronization network model using $|\mathcal{V}_{cut}|+2$ actuation nodes. From Lemma 2, it is immediate that $\mathbf{F}= [\mathbf{\tilde{F}}~~ \mathbf{0}]$ blocks observability in the base synchronization network model. In this way, an observability-blocking controller is found in the base model that uses states of the accessible nodes only. 

The design algorithm requires the graph $\tilde{\mathcal{G}}$ to be strongly connected, and the pair $(\mathbf{\tilde{L}},\mathbf{\tilde{B}})$ to be controllable. The strong connectivity condition guarantees that $\mathbf{\tilde{L}}$ has at least one appropriate eigenvalue, while the controllability condition allows the design of an observability-blocking controller in the accessible-region synchronization network model per Theorem 3. We formalize this result in the following theorem. The proof is omitted as it is immediate from the discussion.

\medskip

\noindent \textbf{Theorem 5:} \textit{Consider a (base) synchronization network model which has a set of accessible nodes as defined above. Say that a separating cutset has been identified, and associated accessible-region synchronization network model has been constructed. Assume that: 1) the number of actuation nodes is two more than the cardinality of the vertex-cutset $\mathcal{V}_{cut}$ ($q=|\mathcal{V}_{cut}|+2$), 2) the graph $\tilde{\mathcal{G}}$ of the accessible-region synchronization network is strongly connected, and 3) the pair $(\mathbf{\tilde{L}},\mathbf{\tilde{B}})$ is controllable. Then the gain matrix $\mathbf{F}$ of the regional state feedback controller, which uses only the states of accessible nodes, can be designed according to the process described above to block observability in the base synchronization network model (i.e. to make the pair $(\mathbf{C},-(\mathbf{L}+\mathbf{BF}))$ unobservable).} 

\medskip

It should be noted that the second condition in Theorem 5 can be relaxed for the special case where inaccessible vertices comprise the partition $\mathcal{V}_2$ entirely. In that case, any eigenvalue of $\mathbf{\tilde{L}}$ can be chosen in the design as $\mathcal{V}_3= \emptyset$, and therefore the graph $\tilde{\mathcal{G}}$ does not need to be strongly connected. This implies that we can also relax this connectivity condition if we consider the accessible-region synchronization network model to be defined on the vertex-set $\mathcal{V}_1 \cup \mathcal{V}_{cut}$ only. However, the feedback control will then be restricted from using the states of the nodes associated to $\mathcal{V}_3$ even though they are accessible.

The design algorithm presented in Theorem 5 has a drawback: the designed controller does not maintain the open-loop eigenvalues of the base synchronization network model. In fact, the closed-loop base model may become unstable (for instance, see Example 2 in Section V). For some special cases, the design can be guaranteed to maintain stability in the closed-loop base model. For instance, it can be argued that the designed regional feedback controller maintains stability in the closed-loop base model when the links between accessible-region synchronization network and inaccessible-region synchronization network are sufficiently weak (or equivalently, weights of the edges between accessible and inaccessible vertices are sufficiently small). However, stability cannot be guaranteed in general for this design.

Now we suggest a modification to the above design algorithm, such that unobservability can be enforced in the base model with a general guarantee on stability. The modified design uses time-scale separation principles. The idea is to design an observability-blocking controller in the accessible-region synchronization network model, which achieves a time-scale separation between the states of accessible and inaccessible nodes.  This is done by shifting the eigenvalues of $\mathbf{\tilde{L}}$ sufficiently far to the right first, and then using the algorithm presented in Section IV.A. For this design, the same argument as for Theorem 5 can be used to verify unobservability, while the time-scale separation guarantees stability. Specifically, the controller is designed as follows. First, as before vertex-cutset $\mathcal{V}_{cut}$ is chosen and the associated accessible-region synchronization network model is formed. Then, using any standard eigenvalue assignment technique (e.g. \cite{pole_1},\cite{pole_2},\cite{pole_3}) the gain matrix $\mathbf{\tilde{F}}_1$ of a state feedback controller is designed so that the closed loop $(\mathbf{\tilde{L}}+\mathbf{\tilde{B}\tilde{F}}_1)$ 
has all eigenvalues with real part greater than a pre-selected nonnegative constant $d$, and further the technical constraint that the assigned eigenvalues do not match with the eigenvalues of the block $\mathbf{\tilde{L}}_{\mathcal{V}_3\mathcal{V}_3}$ is met.   Next, an eigenvalue $\tilde{\lambda}_p$ of $(\mathbf{\tilde{L}}+\mathbf{\tilde{B}\tilde{F}}_1)$ is chosen so that it is not an eigenvalue of block $\mathbf{\tilde{L}}_{\mathcal{V}_3\mathcal{V}_3}$. Thereupon, the observability-blocking controller design algorithm presented in Section IV.A is applied on $(\mathbf{\tilde{L}}+\mathbf{\tilde{B}\tilde{F}}_1)$ to obtain the gain matrix $\mathbf{\tilde{F}}_2$, for which the pair $(\mathbf{\tilde{C}},-(\mathbf{\tilde{L}}+\mathbf{\tilde{B}\tilde{F}}_1+\mathbf{\tilde{B}\tilde{F}}_2))$ has an unobservable mode at $(-\tilde{\lambda}_p)$. The combination of these gains, i.e. $\mathbf{\tilde{F}}=\mathbf{\tilde{F}_1+\tilde{F}_2}$, serves as the observability-blocking controller in the accessible-region synchronization network model. Now according to the Lemma 2, $\mathbf{F}= [\mathbf{\tilde{F}}~~ \mathbf{0}]$ blocks observability in the base synchronization network model. Further, this regional state feedback controller also maintains stability when $d$ is chosen appropriately (e.g. large enough). The following theorem formalizes this.

\medskip

\noindent \textbf{Theorem 6:}  \textit{Consider a (base) synchronization network model which has a set of accessible nodes as defined above. Say that a separating cutset has been identified, and associated accessible-region synchronization network model has been constructed. Assume that: 1) the number of actuation nodes is two more than the cardinality of the vertex-cutset $\mathcal{V}_{cut}$ ($q=|\mathcal{V}_{cut}|+2$), and 2) the pair $(\mathbf{\tilde{L}},\mathbf{\tilde{B}})$ is controllable. Then the gain matrix $\mathbf{F}$ of regional state feedback controller, which uses only the states of accessible nodes, can be designed according to the process described above to block observability in the base synchronization network model (i.e. to make the pair $(\mathbf{C},-(\mathbf{L}+\mathbf{BF}))$ is unobservable) while maintaining stability of the closed-loop system.}

\medskip

\noindent \textbf{Proof:}

For a controllable linear time-invariant system, the eigenvalues can be placed anywhere in complex plane using a state feedback controller \cite{eig-contl_lin}. Since the pair $(\mathbf{\tilde{L}}, \mathbf{\tilde{B}})$ is assumed to be controllable, therefore $\mathbf{\tilde{F}}_1$ as described in the above algorithm can always be found. Subsequently, a second control loop can be applied to block observability, using an argument identical to that for Theorem 5.  Thus, we can conclude that a combined and zero-padded gain $\mathbf{F}= [\mathbf{\tilde{F}}~~ \mathbf{0}]$ can always be found, which first moves the eigenvalues and then blocks observability in the base synchronization network model. 

Now using time-scale separation based argument we will show that there are values of $d$ for which the closed-loop base synchronization network model is stable. Using (\ref{eq7}) and (\ref{eq8}), we can write the closed-loop base model as
\begin{eqnarray} \label{eq9}
(\mathbf{L}+\mathbf{BF})=
\begin{bmatrix}
    (\mathbf{\tilde{L}}+\mathbf{\tilde{B}\tilde{F}})+ \mathbf{P}_1 & \mathbf{L}_{reg,rem}\\
    \mathbf{L}_{rem,reg} & \mathbf{L}_{\mathcal{V}_4\mathcal{V}_4}\\
  \end{bmatrix}
  \end{eqnarray}
where $\mathbf{P}_1$ and $\mathbf{P}_2$ are positive semidefinite and non-zero diagonal matrices, $\mathbf{L}_{reg,rem}= [\mathbf{0}^T~~ \mathbf{L}_{\mathcal{V}_{cut}\mathcal{V}_4}^T ~~ \mathbf{L}_{\mathcal{V}_{3}\mathcal{V}_4}^T]^T$ and $\mathbf{L}_{rem,reg}= [\mathbf{0}~~ \mathbf{L}_{\mathcal{V}_4\mathcal{V}_{cut}} ~~ \mathbf{L}_{\mathcal{V}_4 \mathcal{V}_3}]$. Recall that $\mathbf{\tilde{F}}_2$ maintains the eigenvalues of $(\mathbf{\tilde{L}}+\mathbf{\tilde{B}\tilde{F}}_1)$. Now if we choose $d$ to be very large, then it is obvious that the states of the accessible nodes will have very fast decaying dynamics in the close-loop system. Thereupon by choosing sufficiently large $d$, the states of the accessible nodes can be made separated in time scale from those of inaccessible nodes. In that case the stability of the closed-loop model can be addressed via singular perturbation theory. Specifically in the case of such time-scale separation, according to the singular perturbation theory \cite{sp_1} the closed-loop system of the base synchronization network model will be stable, if $\mbox{Re}\{\lambda\{\mathbf{\tilde{L}}+\mathbf{\tilde{B}\tilde{F}}+ \mathbf{P}_1 \}\} > 0 $ and $\mbox{Re}\{ \lambda\{\mathbf{L}_{\mathcal{V}_4\mathcal{V}_4}- \mathbf{L}_{reg,rem}~ {(\mathbf{\tilde{L}}+\mathbf{\tilde{B}\tilde{F}}+ \mathbf{P}_1)}^{-1}~ \mathbf{L}_{rem,reg}\}\} > 0$. (Here, $\mbox{Re}\{\lambda\{\mathbf{A}\}\}$ denotes the real parts of all the eigenvalues of $\mathbf{A}$ following the notation in \cite{sp_1}.) Now for arbitrarily large $d$ these conditions reduce to $\mbox{Re}\{\lambda\{\mathbf{L}_{\mathcal{V}_4\mathcal{V}_4}\}\} > 0$ as $(\mathbf{\tilde{L}}+\mathbf{\tilde{B}\tilde{F}}+ \mathbf{P}_1)^{-1} \to \mathbf{0}$. The proof is completed by noting that $\mathbf{L}_{\mathcal{V}_4\mathcal{V}_4}$ is a grounded Laplacian in a strongly connected graph and thus all of its eigenvalues are in the open right half plane \cite{gl_1}. \hfill $\blacksquare$

\medskip

{\em Remark:} When all the eigenvalues of $\mathbf{L}$ are real, we can use one fewer actuation node  ($q=|\mathcal{V}_{cut}|+1$ actuation nodes) in our design.

Theorem 6 suggests that we can design observability-blocking controller which uses states of the accessible nodes only and preserves stability as long as the pair $(\mathbf{\tilde{L}},\mathbf{\tilde{B}})$ is controllable and there are sufficient number of actuation nodes.  Theoretically, an arbitrarily large value of $d$ in the design leads to arbitrarily large gains in the controllers. However, in practice arbitrarily large $d$ is not necessary in the design to achieve stability, as the edge-weights in the real-world networks are finite and in many cases small. Therefore for a real-world network satisfying the conditions given in Theorem 6, it should be possible to design regional state feedback observability-blocking controller with a relatively limited control gain that maintains stability (by choosing a moderate value of $d$). 

Many real-world networks e.g. power,  communication, and traffic networks,  are sparse. The regional feedback controller design algorithm may be useful for such networks, in that it can permit authorities to prevent learning of network dynamics even without full measurements of the network state. As a limiting case, if the network graph is tree, observability can be blocked for a connected subnetwork using only two actuators, and without any access to the subnetwork states, assuming that the accessible-region model is controllable.

\section{Numerical Examples}

Two numerical examples are presented to illustrate the sparser controller design and the regional state feedback controller design, respectively. 

\begin{figure}[thpb]
\centering
\includegraphics[width=8cm,height=4cm]{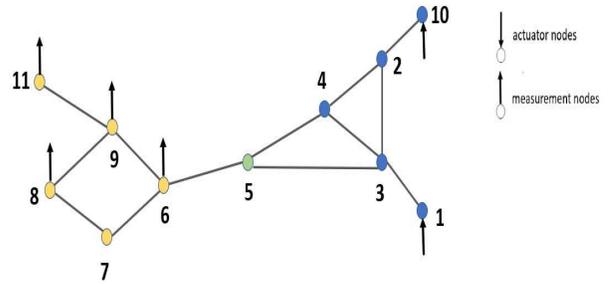}
\caption{Network graph for Example 1.}
\end{figure} \label{fig4}

\noindent \textbf{Example 1}: The sparser observability-blocking controller, as formalized in Theorem 4, is illustrated. A network with $11$ nodes is considered. The network's graph is shown in Fig. 3: all edges shown in the figure are bi-directional, and the edge-weights are set to $1$.  Nodes $\{ 6, 8, 9, 11 \}$ are measurement nodes, while Nodes $\{1, 2, 3, 4, 5, 10\}$ are actuation nodes. 

Our goal is to design a sparse state feedback controller that blocks observability at these measurement nodes. In this example, Vertex 5 is a single vertex-cutset that separates the measurement and possible actuation vertices in the network graph. Thus, according to Theorem 4, we should be able to block observability at the measurement nodes by blocking observability at Node 5, using two actuation nodes. (Note that the graph is undirected, therefore $q=|\mathcal{V}_{cut}|+1$ actuation nodes suffices.) We select Nodes 1 and 10 as the actuation nodes.  The model is controllable for this selection. 

We apply the controller design process described in Section IV.B, which uses Algorithm 1, to design the feedback controller.  Specifically, we choose $\lambda_p=0.1853$ from the eigenvalues of $\mathbf{L}$. (It is easy to check that $\lambda_p$ is not an eigenvalue of block $\mathbf{L}_{\mathcal{V}_2\mathcal{V}_2}$.) Then following the algorithm, the control gains for controllers at Nodes 1 and 10 are obtained as $[0.7311;$ $0.7281;$ $0.5956;$ $0.5601;$ $ 0.2527;$ $-0.4445;$ $-0.7016;$ $-0.8286;$ $-0.8021;$ $0.8938;$ $-0.9846]$ and  $[-0.5980;$ $-0.5956;$ $-0.4872;$ $-0.4581;$ $-0.2067;$ $ 0.3636;$ $0.5739;$  $0.6778;$ $0.6561;$ $-0.7311;$ $0.8054]$, respectively. 

In the closed-loop system, the eigenvector associated with $\lambda_p$ is modified to $[-0.6008;$ $0.2656; -0.0696;$ $0.0696;$ $0; 0; 0; 0;$ $0; 0.7475; 0]$, while other eigenvectors and all the eigenvalues remain unchanged. The zero entries of this modified eigenvector indicate that that the base model has been made unobservable, using only two controllers as dictated by Theorem 4. In fact the network dynamics is now unobservable no matter which nodes in the partition $\{5, 6, 7, 8, 9, 11\}$ are measured.

\begin{figure}[thpb]
\centering
\includegraphics[width=8cm,height=4cm]{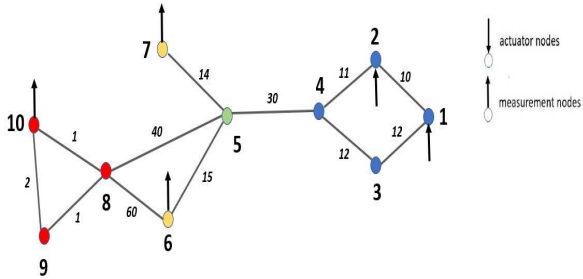}
\caption{Network graph for Example 2}
\end{figure} \label{fig3}

\noindent \textbf{Example 2}: A regional state feedback control design which blocks observability while maintaining stability via a time-scale separation is illustrated. A network with $10$ nodes is considered. The network's graph is shown in Fig. 4: all edges shown in the figure are bi-directional, and the number shown next to each edge is the weight of that edge. The measurement nodes are Nodes $\{ 6,7,10 \}$, while the actuation nodes are Nodes $\{1, 2, 3, 4, 5\}$. In this example we further assume that only the states of the Nodes $\{1,2,3,4,5,6,7\}$ are accessible to the control authorities. Our goal is to design an observability-blocking controller that only uses the states of these accessible nodes. Here, we have deliberately constructed the network so that a naive design using Theorem 5 leads to instability, and hence a time-scale separation-based design is needed.  

\noindent  

In this example, Vertex $5$ is a single vertex-cutset which partitions the network graph as required for the regional controller design: specifically, it forms a partition with only accessible vertices, which contains the possible actuation vertices and no measurement vertices. The design depends on the construction of accessible-region synchronization network model on Nodes $\{1,2,3,4,5,6,7\}$ with the measurement node as Node $5$ and two actuation nodes, as described in Section IV.C. (The graph is undirected, hence $q=|\mathcal{V}_{cut}|+1$ actuation nodes suffices.) We select Nodes 1 and 2 as the actuator nodes; the accessible-region synchronization model can be checked to be controllable. The state matrix for the synchronization dynamics of this model is given by the following Laplacian matrix $\mathbf{\tilde{L}}$.
\begin{equation}
\tiny
\setcounter{MaxMatrixCols}{15}
\setlength\arraycolsep{4pt}
\mathbf{\tilde{L}}=\begin{bmatrix*}[r]
22 & -10 & -12 & 0 & 0 & 0 & 0 \\
-10 & 21 & 0 & -11 & 0 & 0 & 0 \\
-12 & 0 & 24 & -12 & 0 & 0 & 0 \\
0 & -11 & -12 & 53 & -30 & 0 & 0\\
0 &  0 & 0 & -30 & 59 & -15 & -14 \\
0 & 0 & 0 & 0 & -15  & 15 & 0\\
0 & 0 & 0 & 0 & -14  & 0 & 14\\
\end{bmatrix*} \nonumber
\end{equation}

We first design the regional state feedback observability-blocking controller based on Theorem 5. For the design, we choose the eigenvalue $\tilde{\lambda}_p= 14.4812$ of $\mathbf{\tilde{L}}$, and then apply Algorithm 1 to obtain the gain matrix $\mathbf{\tilde{F}}$ that blocks observability in the  accessible-region synchronization network model. Per Theorem 5, the gain $\mathbf{F}= [\mathbf{\tilde{F}} ~ \mathbf{0}]$ then blocks observability in the base model. However, for this specially constructed example, the closed-loop system has a pair of unstable modes ($0.0548 \pm 2.8411i$). 

Since the naive design does not preserve stability, we use the alternate stability preserving design. To do so, we first use feedback to shift all eigenvalues of the Laplacian $\mathbf{\tilde{L}}$ to have real parts greater than $10$, using a standard eigenvalue placement technique. (In fact, we only place eigenvalues at $0$ and $6.5855$ at $11$ and $12$ respectively keeping the remaining eigenvalues fixed.) Then we choose the same $\tilde{\lambda}_p$ as before, and use the developed algorithm to find the gain matrix $\mathbf{\tilde{F}}$ that blocks observability in the  accessible-region synchronization network model. We notice that $\mathbf{F}= [\mathbf{\tilde{F}} ~ \mathbf{0}]$ blocks observability in the base model and also maintains closed-loop stability. Here, the control gains for the controllers at Node 1 and 2 are given by $[0.8659;$ $-5.0798;$ $-11.1017;$ $79.5732;$ $-91.3091;$ $13.5638;$ $12.1537;$ $0; 0; 0]$ and $[22.6071;$ $15.6908;$ $14.2283;$ $51.1514;$ $1.6000;$ $45.4956;$ $45.3810;$ $0; 0; 0]$. It should be noted that the last 3 entries of the controller gains of the both controllers are zero, and thus the states of the inaccessible nodes, i.e. Nodes $\{ 8, 9, 10 \}$, are not used by the feedback controller.

\section{Conclusions and Future work}

Designs  for control systems in network synchronization processes are presented, which prevent observers at remote nodes from having full visibility of the state dynamics.  These designs exploit eigenstructure assignment techniques to block observability, while also preserving much of the system's eigenstucture.  Sparser control schemes that exploit the network's graph topology, as well as feedback schemes which only require regional data, have also been developed.  These sorts of design algorithms may be valuable for network-control applications where security and privacy needs are paramount, since they can allow operators to keep adversaries and other stakeholders from having full visibility of the dynamics.  From a methodological standpoint, our study begins to address design in networks with multiple orthogonal control authorities, by showing how one controller can be designed to shape remote channels or input-output processes.   

A number of directions of further work are of interest including: 1) generalizing the algorithm to more complex network models, 2) achieving sparser designs by allowing some further flexibility in the eigenvalue/eigenvector placements, and 3) developing algorithms that block particular critical statistics from being estimated.

\subsection*{Acknowledgements}

The authors gratefully acknowledge the support of the United States National Science
Foundation under grants CNS-1545104 and CMMI-1635184.

\end{document}